\documentclass[twocolumn,pre,aps,showpacs,a4paper,floatfix]{revtex4-1}
\pdfoutput=1
\usepackage{graphicx}
\usepackage{psfrag}
\usepackage{subfigure}
\usepackage{epsfig}
\usepackage{calc}
\usepackage{bm}
\usepackage{longtable}
\usepackage{amsmath}
\usepackage{amssymb}
\usepackage[utf8]{inputenc}
\usepackage{color}

\DeclareUnicodeCharacter{2014}{\dash}

\newcommand{\kappabar}{\bar{\kappa}}

\newcommand{\rhobf}{\bm{\rho}}

\newcommand{\beq}{\begin{equation}}\newcommand{\eeq}{\end{equation}}\newcommand{\beqa}{\begin{eqnarray}}\newcommand{\eeqa}{\end{eqnarray}}\newcommand{\boldnabla}{\mbox{\boldmath$\nabla$}}

\begin{document}

\title{ Wrapping of ellipsoidal nano-particles by fluid membranes}
\author{
Sabyasachi Dasgupta, Thorsten Auth and Gerhard Gompper
 }
\affiliation{Theoretical and Soft Matter Biophysics, Institute of Complex Systems and
 Institute for Advanced Simulations, Forschungszentrum J\"ulich, D-52425 J\"ulich, Germany}

\begin{abstract}
Membrane budding and wrapping of particles, such as viruses and nano-particles, play a key role in intracellular transport and have been studied for a variety of biological and soft matter systems. We study nano-particle wrapping by numerical minimization of bending, surface tension, and adhesion energies. We calculate deformation and adhesion energies as a function of membrane elastic parameters and adhesion strength to obtain wrapping diagrams. We predict unwrapped, partially-wrapped, and completely-wrapped states for prolate and oblate ellipsoids for various aspect ratios and particle sizes. In contrast to spherical particles, where partially-wrapped states exist only for finite surface tensions, partially-wrapped states for ellipsoids occur already for tensionless membranes. In addition, the partially-wrapped states are long-lived, because of an increased energy cost for wrapping of the highly-curved tips. Our results suggest a  lower uptake rate of ellipsoidal particles by cells and thereby a higher virulence of tubular viruses compared with icosahedral viruses, as well as co-operative budding of ellipsoidal particles on membranes.
\end{abstract}

\maketitle

\section{Introduction}
 
\footnotetext{\textit{$^{\ast}$~Theoretical Soft Matter and Biophysics, Institute of Complex Systems and Institute for Advanced Simulations, Forschungszentrum J\"ulich,
D-52425 J\"ulich, Germany. E-mail: s.dasgupta@fz-juelich.de; t.auth@fz-juelich.de; g.gompper@fz-juelich.de }}

Budding and subsequent vesiculation of lipid bilayer membranes is essential for transport in biological cells 
\cite{mcmahon05,Hurley_2010,Canton_Battaglia_2012}. Biomembranes usually consist of a variety of lipids and proteins, therefore buds can be induced by lipid domain formation \cite{sunilkumar99,sunilkumar01,baumgart03,semrau08}, by membrane spontaneous curvature
\cite{Kohyama_Kroll_Gompper03,reynwar07,Bassereau_2007,sens08,auth09},  or by a combination of both \cite{Lipowsky92,Lipowsky93,Juelicher93}. For particle wrapping, in addition to the membrane properties, the particle shape and the adhesion strength have to be taken into account \cite{Champion_Mitragotri06,Chithrani06,Decuzzi_Ferrari_2009,Frank12,Florez_landfester12}. Biological examples are viral budding \cite{tzlil04,Mercer_Helenius_2010} and --- for designing efficient drug delivery systems and other nano-engineered techniques for medical diagnosis --- the uptake of small particles by cells  \cite{Chithrani06, Gratton_DeSimone_2008,Florez_landfester12,barua13}. Elongated viruses have been found to form patterns on cell membranes \cite{Kubo_2009}, and ellipsoidal nano-particles are used for drug delivery \cite{liu2012} and as markers \cite{Pissuwan_Cortie_2008,Xu_2011} in cell biology.

Techniques to fabricate nano-particles of different shapes and sizes are well established, thus wrapping has been studied experimentally using both vesicles and cells. To gain insight into the basic mechanism of cellular uptake, the role of shape and size has been investigated both in vitro \cite{Champion_Mitragotri06,Chithrani06,Gratton_DeSimone_2008, Florez_landfester12} and as well as in vivo \cite{Dietrich_Angelova_Pouligny_1997,Lipowsky_Dobereiner_1998,Zhang_Beales_2012} studies.  
 

Describing the membranes as a mathematical surface with appropriate curvature-elastic constants, continuum models predict wrapping as function of bending rigidity $\kappa$, spontaneous curvature $c_0$, surface tension $\sigma$, and for lipid domain formation also the line tension $\gamma$ at the domain boundary. Bending rigidity and surface tension oppose wrapping, whereas the adhesion strength $w$ favors wrapping. For example, for a completely wrapped sphere of radius $R$ and for a lipid bilayer without surface tension, an adhesion energy gain of $-4 \pi R^2 w$ is opposed by a bending energy cost of $8 \pi \kappa$. Complete wrapping occurs if the adhesion strength exceeds $w^* = 2 \kappa/R^2$, while for smaller values the sphere remains unwrapped.
 
Wrapping of spherical particles has been studied systematically using continuum membranes and is well understood. A particle wrapped by an infinitely large planar membrane without surface tension is fully described by the simple calculation above. Neglecting a surface or volume constraint, also the energy of a vesicle that wraps a particle has been obtained analytically \cite{Benoit_Saxena_2007}. For membranes with surface tension, the deformation energy of the membrane can be calculated using approximate models \cite{Dietrich_Angelova_Pouligny_1997,deserno02} or shape equations that are evaluated numerically \cite{deserno03,deserno04,deserno04a}. Similarly, the deformation energy for wrapping of infinitely long cylinders has been calculated \cite{Weikl03, Mkrtchyan_Chen10}, which is qualitatively different from wrapping of spheres. Regarding the example given above, the free membrane around the sphere forms a catenoid without any bending energy cost, whereas for a cylinder wrapped by an infinite planar membrane, the deformation energy of the free membrane has to be
taken into account.
 
In this article, we investigate wrapping of ellipsoidal particles by homogeneous membranes without spontaneous curvature. In section \ref{sec3a}, we calculate the wrapping energy as function of the wrapping fraction for membranes with and without surface tension. We obtain phase diagrams that show a non-wrapped state, a partially-wrapped state, and a fully-wrapped state, see section \ref{sec3b}. While the transition between the unwrapped and the partially wrapped state is continuous, the partially-wrapped state is separated from the fully-wrapped state by an energy barrier. In section \ref{sec3c}, we characterize the energy barrier and the hysteresis that is found for the transition between the partially-wrapped and the fully-wrapped state. Finally, in section \ref{sec3d} we discuss the role of the shape for the wrapping energy.

\section{Theoretical Model and Methods}
\label{sec2}

\subsection{Continuum membrane model}
\label{continuummodel}

\begin{figure}[t]
  \begin{center}
    \leavevmode
    \includegraphics[width=0.80\columnwidth]{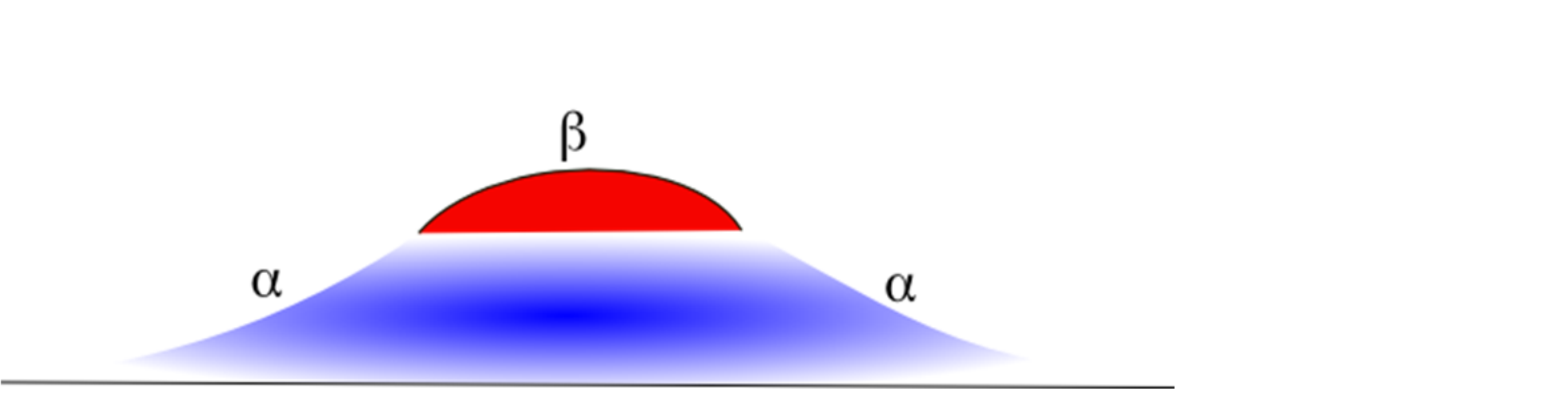}
    \hspace{5ex}
   \end{center}
    \caption{Deformation profile of membrane adhering to a rigid nano-particle. The  free membrane around the particle is labeled $\alpha$ and the membrane adhered to the particle $\beta$.}
    \label{MPWC}
\end{figure}

Using the continuum membrane model, our system is constructed in order to include the minimal ingredients required to characterize wrapping of ellipsoidal particles. The uptake process can be understood as a competitive tug of war between the elastic deformation energy and the contact interaction between particle and membrane. The elastic deformation energy $\mathcal{ E_{\textrm{def }}}$ of the lipid bilayer is \cite{Canham1970,Helfrich_1973}
\begin{equation}
\mathcal{ E_{\textrm{def }}} = \int_{A_{\rm tot}} d S \, \left[ 2 \kappa (H - c_0)^2 + \kappabar K  +  \sigma   \right] \, ,
\end{equation}
which is an integral over the entire membrane surface area, $A_{\rm tot}$. The membrane shape is described by the mean curvature, $H=(c_1+c_2)/2$, and the Gaussian curvature, $K=c_1 c_2$;  $c_1$ and $c_2$ are the principal curvatures of the membrane.  The surface tension $\sigma$ is the Lagrange multiplier conjugate to the excess area that is generated due to wrapping on the particle relative to the unwrapped flat membrane. Because we assume a symmetric lipid bilayer, we use $c_{0} = 0$. The integral over the Gaussian curvature with the constant saddle splay modulus $\kappabar$ is determined by the topology of the membrane (and by the geodesic curvature at a boundary). In our case of an infinite planar membrane, the integral is a constant during the wrapping process. Thus, the total energy for a membrane-particle  wrapping complex is 
\begin{equation}
\mathcal{ E_{\textrm{tot}}} = \int_{A_{\rm tot}} d S \, \left[  2 \kappa  H  ^2      +  \sigma \right]  - w \int_{A_{\rm ad}} d S \, .
\label{maineq}
\end{equation}
The contact interaction with adhesion strength $w$ is proportional to the membrane area $A_{\rm ad}$ adhered to the particle, see Fig.~\ref{MPWC}.

Our continuum model in Eq.~(\ref{maineq}) is applicable for particle sizes that are larger than a few times the thickness of a  lipid bilayer, which is about $ 5 \, {\rm nm}$. For particle sizes smaller than $\sqrt{\kappa/\sigma}$ both bending energy and surface tension contribute, for larger particle sizes surface tension is dominant \cite{deserno04}. A characteristic length scale for each system is the
particle size $a$, typically $20-100 \, \rm nm$ and a characteristic energy scale is the bending rigidity $\kappa$, typically $10-100 \, k_{\mathrm{B}}T$. We therefore describe
our system in terms of dimensionless parameters, which we choose to be consistent with those in Refs.~\onlinecite{deserno04,deserno04a}. This gives 
\begin{equation}
\tilde {E}  = \frac {1}{2 \pi a^2 } \left( \int_{A_{\rm tot}} d S \, \left[  4  (aH)  ^2      +  2 \tilde{\sigma} \right]  - \tilde {w} \int_{A_{\rm ad}} d S  \right) \, ,
\end{equation}
with the reduced energy $\tilde{ E } = \mathcal{ E_{\rm tot}}/(\pi \kappa)$, the reduced surface tension $\tilde{\sigma} = \sigma a^2/\kappa$, and the reduced adhesion strength $\tilde{w} = 2w a^2/\kappa$.

\subsection{Energy minimization}

There are three approaches to calculate the membrane deformation for particle wrapping \cite {Bloor_Wilson00}. (i) Solving the Euler-Lagrange equation: the Euler-Lagrange equation is obtained from Eq.~(\ref{maineq}) using variational calculus; it is a fourth-order partial non-linear differential equation \cite{OuYang_Helfrich_1989} and a general solution does not exist. Exploiting symmetry and choosing an appropriate parametrization \cite{Helfrich_1973, Seifert_Berndl_1991,Julicher_Lipowsky96} , such as the arc-length parametrization proposed in Ref.~\onlinecite{ Julicher_Lipowsky96}, one obtains a set of second order non-linear ordinary differential equations. This method has been employed both for spherical particles in Refs.~\onlinecite{deserno03,deserno04,deserno04a,nowak08} and for infinite cylinders in  Refs.~\onlinecite{nowak08,Mkrtchyan_Chen10}. In the weak-wrapping limit, the Monge parametrization and the approximation for almost planar membranes can be used \cite{Weikl03,deserno03,deserno04,deserno04a}.
(ii) Exploiting a variational approach: minimum energy shapes are found using appropriate functional parametrization, e.~g.\ spherical harmonics \cite{Khairy_Howard11}, Cassini ovals \cite{Canham1970}, and Fourier functions \cite{gozdz01}. The set of parameters obtained from energy minimization describes the membrane deformation. This method has been used for
wrapping in Ref.~\onlinecite{Gozdz_2007} and can treat non-axisymmetric shapes as well. (iii) Using triangulated membranes: the method offers a high flexibility and has been used to study both minimal energy shapes\cite{Brakke92,Wintz_Dobereiner_Seifert_1996} and systems at finite temperature\cite{kantor1989,gompper92,gompper1997,fosnaric09}. The membrane is constructed by small triangles and discretized expressions for the deformation and interaction energies are used. Triangulated membranes provide a huge amount of freedom regarding the choice of shape and local properties of the membrane and has been used to study wrapping of single as well as multiple particles \cite{fosnaric09,Cacciuto_2012a, Bahrami_2012}. We employ ``Surface Evolver'' \cite{Brakke92} for our calculations, a finite-element code for surfaces formed of vertices, edges and facets; the discretization of the bending energy is discussed in appendix \ref{A2}.

\subsection{Wrapping energy calculations}

\begin{figure}[t]
\begin{center}$
\begin{array}{cc}
\includegraphics[trim= 0cm 0cm 0cm 0cm, clip=true, totalheight=0.42\columnwidth]{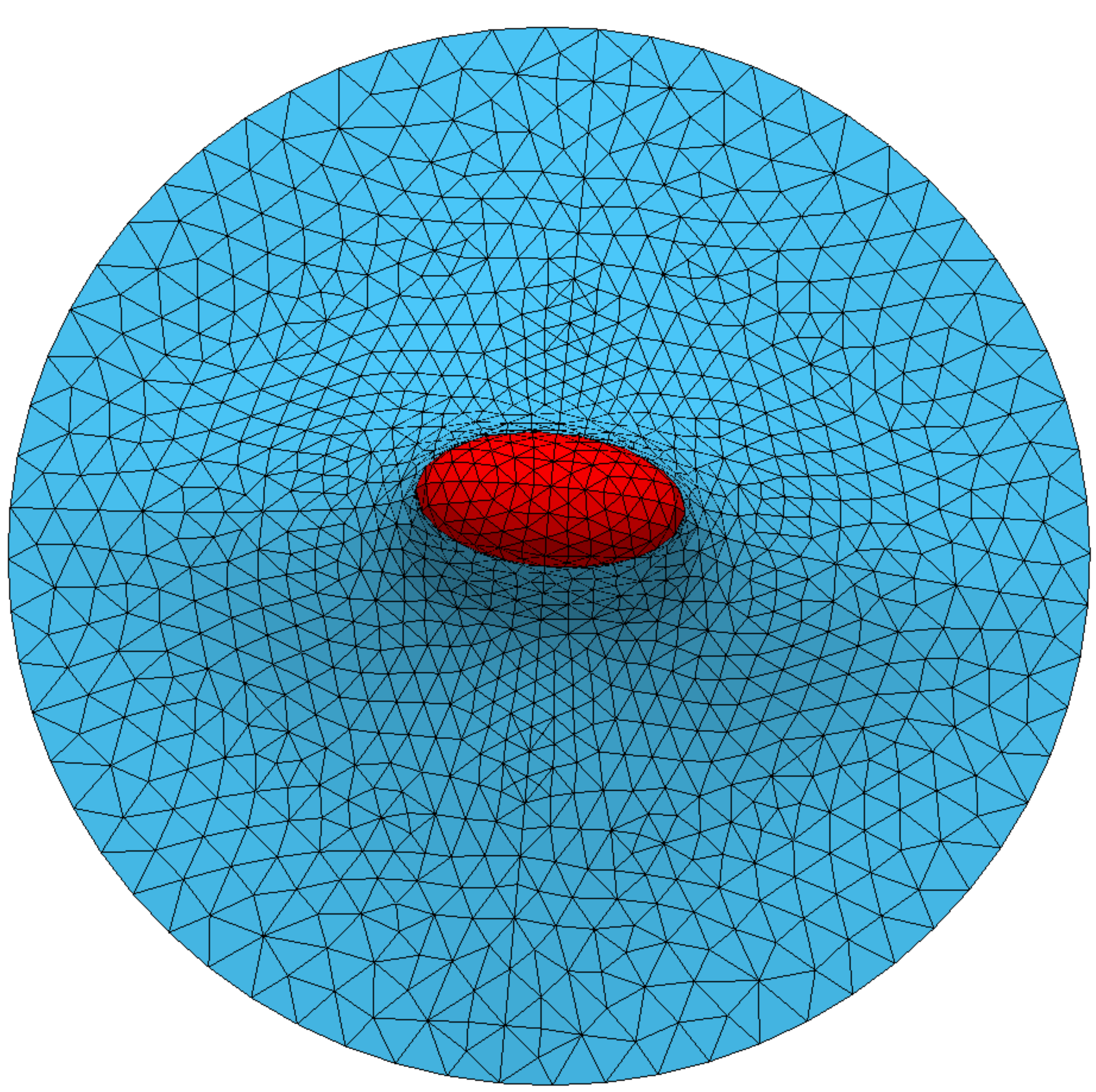} &
\includegraphics[trim= 0cm 0cm 0cm 0cm, clip=true, totalheight=0.21\columnwidth]{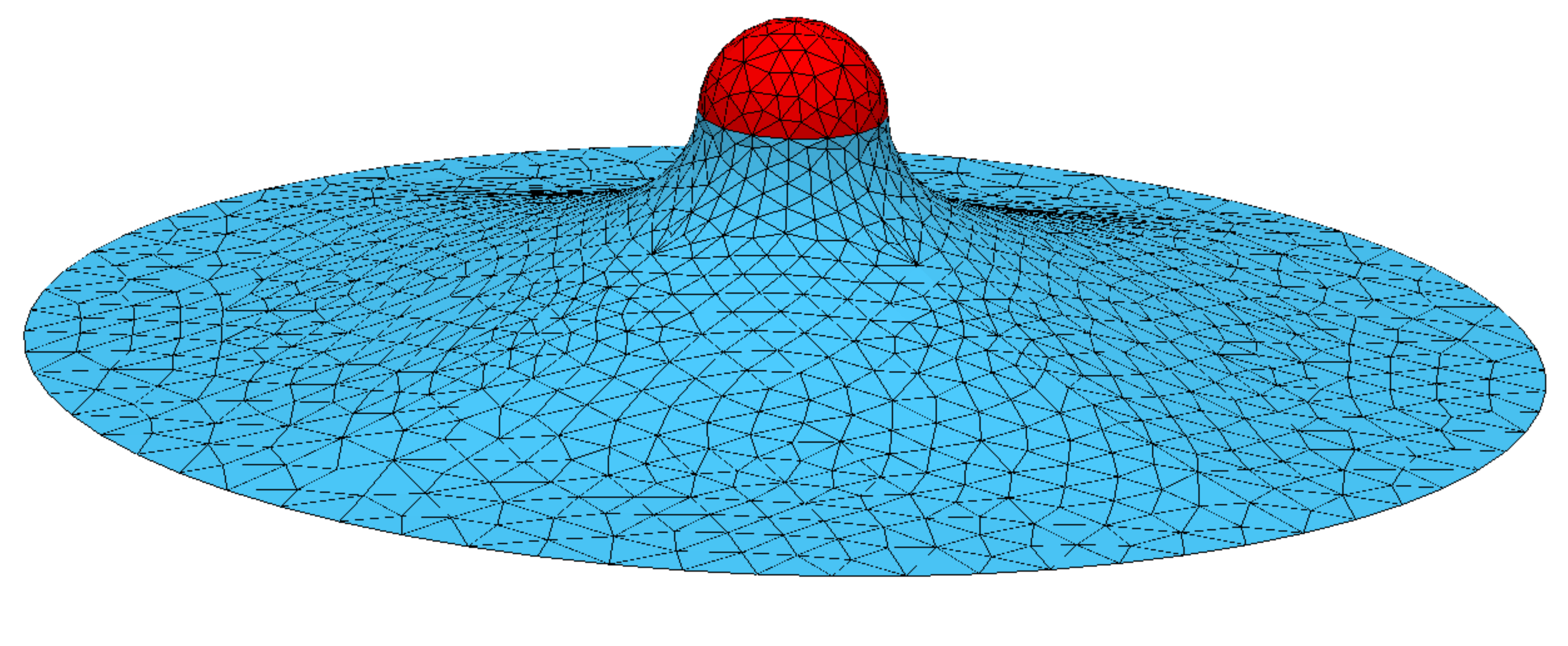}
\end{array} $
\end{center}
\caption{Membrane deformation for membrane-particle systems that have been discretized using triangulated surfaces.}
\label{TRI}
\end{figure}

Deformation energies are calculated such that the membrane wraps the particle starting from the lowest curvature region progressively. Therefore, ellipsoidal particles are oriented with their major axis aligned parallel to a membrane patch that is enclosed by a circular wire frame of radius $20 \, a$, see Fig.~\ref{TRI}. This orientation is energetically favourable until about half wrapping compared with the perpendicular orientation. It ensures a maximum gain of adhesion energy and a minimal cost due to bending and surface tension contributions. We characterize the ellipsoids by their polar radius $a$ and their aspect ratio $b/a$. Prolate ellipsoids correspond to $b/a > 1$ and oblate ellipsoids correspond to $b/a < 1$.

From numerical calculations for ellipsoids with parallel orientation, we obtain the deformation energy as function of the fraction of the particle area that is wrapped, $A_{\rm ad}/A$
\footnote{The wrapping fraction is defined as the ratio of the particle area adhered  to the membrane, $A_{ad}$, to the total area of the particle, $A$. $A_{ad}/A=0 $ corresponds to the unwrapped state, while $A_{ad}/A \approx 1 $ indicates a fully-wrapped state.}. Snaphots of partially-wrapped particles with different wrapping fractions are shown in Figs.~\ref{WEDS}~(b), \ref{WEDPEOE}~(b), and \ref{WEDPEOE}~(d) \footnote{Unlike for prolate ellipsoids at a phase boundary dominated by interfacial tension \cite{lehle08}, and although the boundary between the adhered and the free membrane is allowed to adjust freely in early stages of the energy minimization, we find an elliptical contact line in Fig.~\ref{WEDPEOE}~(b) without significant variation in its height.}. The energy of the membrane relative to the unwrapped state is denoted by $\Delta E$, which gives the wrapping energy cost $\Delta \tilde{E} = \Delta E / \pi\kappa$ in reduced units. To calculate small energy differences and derivatives with sufficient accuracy, this deformation energy profile is fit to the generalized logistic function,
\small
\begin{equation}
f\left(\frac{A_{ad}}{A}\right) = c_{1} \left[ \left( 1 + \exp \frac{ c_{2} - A_{\rm ad}/A}{c_{3}} \right)^{-1} - \left( 1 + \exp \frac{c_{2}}{c_{3}}\right)^{-1} \right]   \, ,
\label{fiteq}
\end{equation}
with the fit parameters $c_{1}$, $c_{2}$, and $c_{3}$. This monotonic function describes the numerical data very well and vanishes for a completely detached particle, see Figs.~\ref{WEDS} and \ref{WEDPEOE}. However, it cannot capture the partially-wrapped state encountered during the unwrapping transition at almost complete wrapping. This high wrapping fraction partially-wrapped state that has been found in Ref.~\onlinecite{deserno04} is a very shallow energy minimum, which we cannot identify in our numerical calculations. Further analysis is done using the fit function, in particular the wrapping energy at any reduced adhesion strength is
 \begin{equation}
 \frac{\Delta E(A_{\rm ad}/A,w)}{ \pi \kappa} = f \left(\frac{A_{\rm ad}}{A} \right) - \tilde {w} \frac{A_{\rm ad}}{A} \, .
\label{fiteqw}
\end{equation} 
Figs.~\ref{WEDS}~(a), \ref{WEDPEOE}~(a), and \ref{WEDPEOE}~(c) show energies as function of the wrapping fraction at reduced tension $\tilde \sigma = 1$ for wrapping of spherical, prolate, and oblate particles respectively. The open circles are the numerically calculated deformation energies that are fit by the solid line given by Eq.~(\ref{fiteq}). The deformation energy thus increases monotonic with the wrapping fraction, the unwrapped state is stable. For finite adhesion strengths, the wrapping energies that are calculated using Eq.~(\ref{fiteqw}) are non-monotonic functions of the wrapping fraction, such that partially and fully-wrapped states can be the stable states.
 
\section{ Results }
\label{sec3}

\subsection{Wrapping energy}
\label{sec3a}
 
 \begin{figure}  [!htb]
\centering
\begin{tabular}{cc}
\raisebox{6.4cm}{(a)} & \includegraphics[trim= 1.5cm 0.1cm 0.5cm 0.1cm, clip=true, totalheight=0.8\columnwidth]{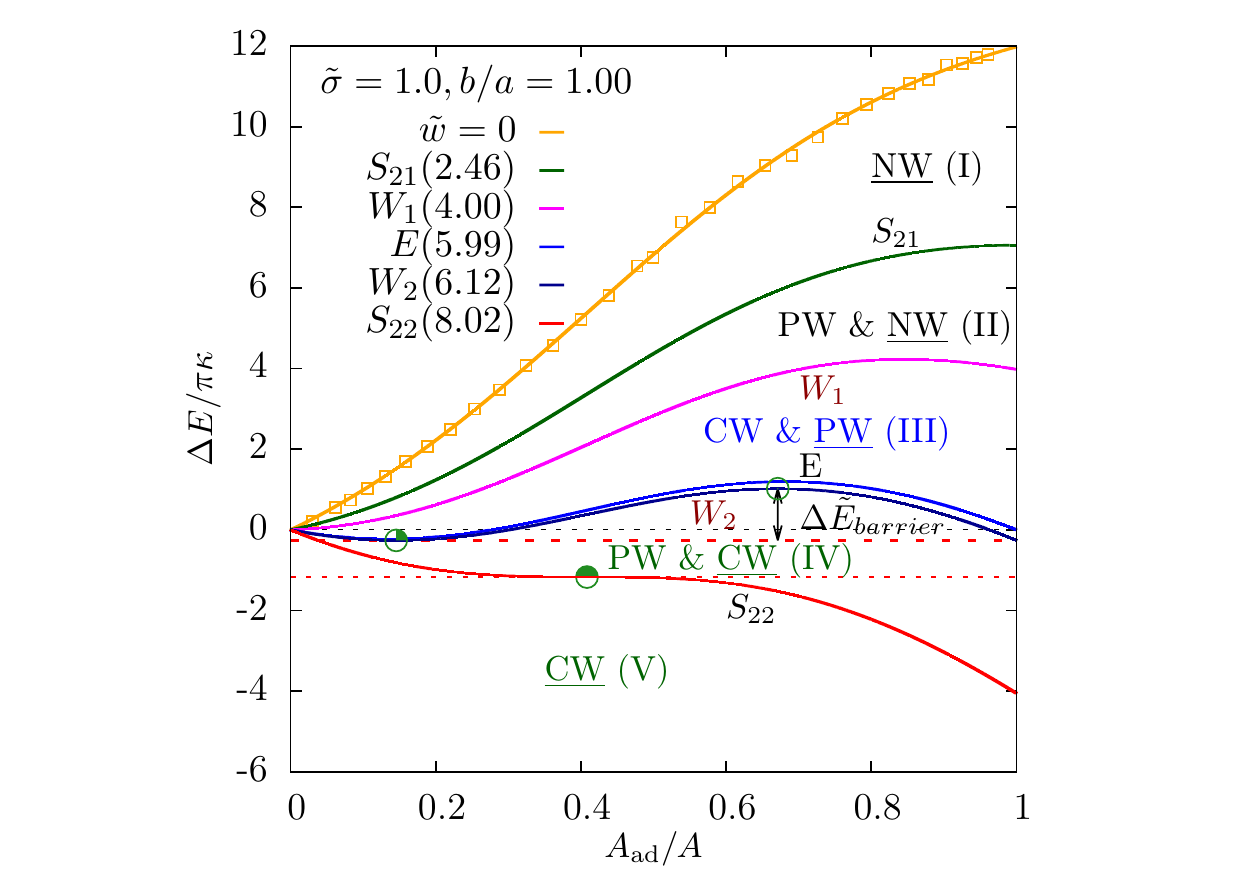}  \\[2ex]
\raisebox{1.4cm}{(b)} & \includegraphics[trim= 0cm 0cm 0cm 0cm, clip=true, totalheight=0.4\columnwidth , angle=90]{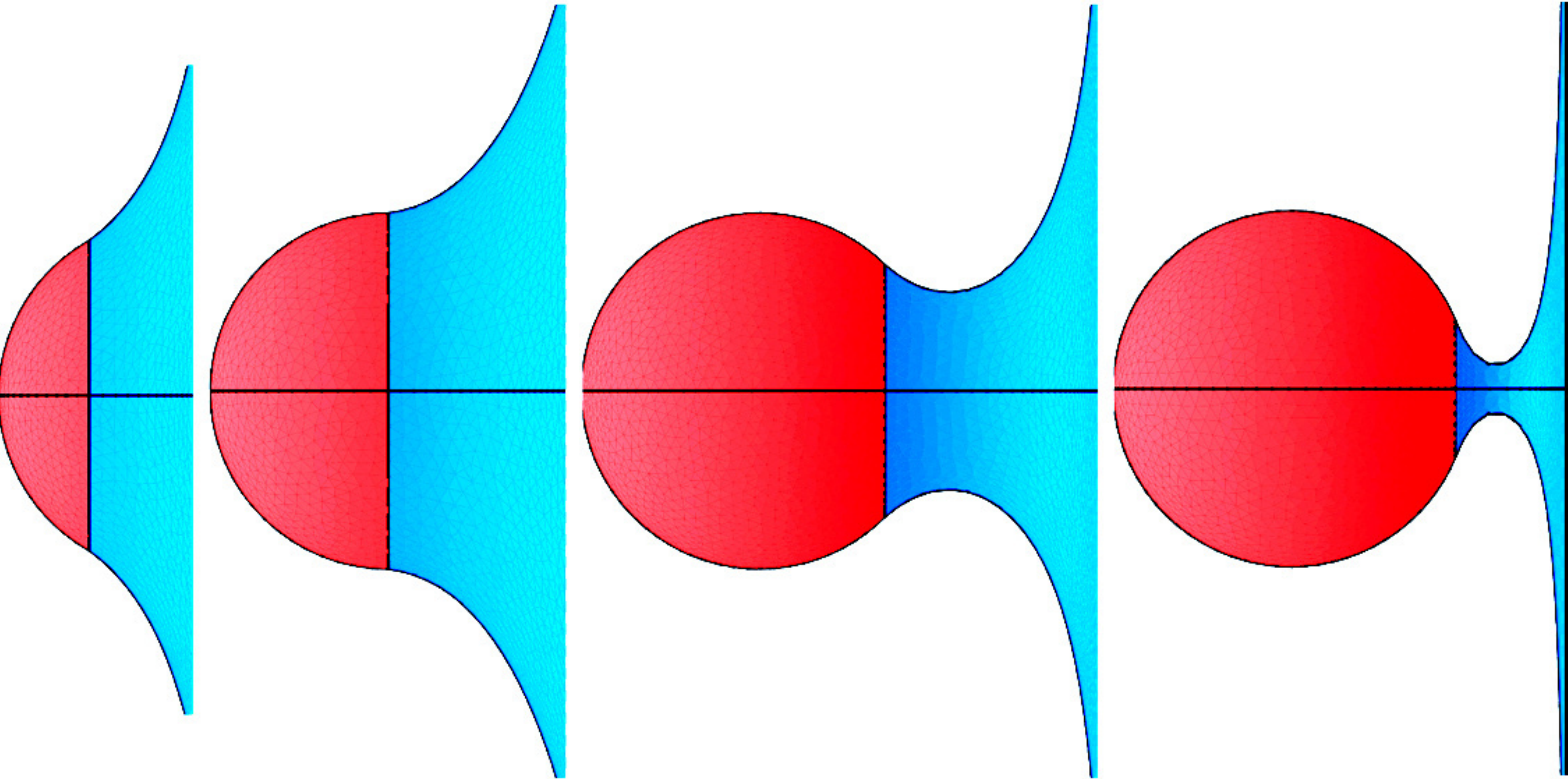}
\end{tabular}
\caption{(a) Energies for wrapping a spherical particle as function of the wrapping fraction ${ A_{\rm ad}}/{A}$ for reduced membrane tension $\tilde \sigma = 1$. The figure shows the wrapping energy profiles for six adhesion strengths: the numerically calculated data for zero adhesion strength and the corresponding fit function, E with equal energy for the non-wrapped and the completely-wrapped states ($\tilde{w}=5.99$), the binding transition ${W_\mathrm{1}}$ between the unwrapped and the partially-wrapped ($\tilde{w} = 4.00$), the binodal ${W_\mathrm{2}}$ between the partially-wrapped and the completely-wrapped state ($\tilde{w} = 6.12$), and the spinodals ${S_\mathrm{21}}$ and ${S_\mathrm{22}}$ that are associated with ${W_\mathrm{2}}$ (for $\tilde{w} = 2.46$ and $\tilde{w} = 8.02$ respectively). The phase boundaries separate $5$ regimes in the phase diagram with stable and metastable completely wrapped (CW), partially-wrapped (PW), and and non-wrapped states (NW); the stable state is underlined. The wrapping fractions that are plotted in Fig.~\ref{PWF} are marked by circles and the energy barrier shown in Fig.~\ref{EBAR} is indicated. (b) Sketches for spherical particles for wrapping fractions 0.25, 0.50, 0.85 and 0.96, with the adhered membrane in red and part of the free membrane in blue. }
\label{WEDS}
\end{figure}

\begin{figure}  [!htb]
\centering
\begin{tabular}{cc}
\raisebox{6.4cm}{(a)} & \includegraphics[trim= 1.8cm 0.1cm 2.2cm 0cm, clip=true, totalheight=0.8\columnwidth]{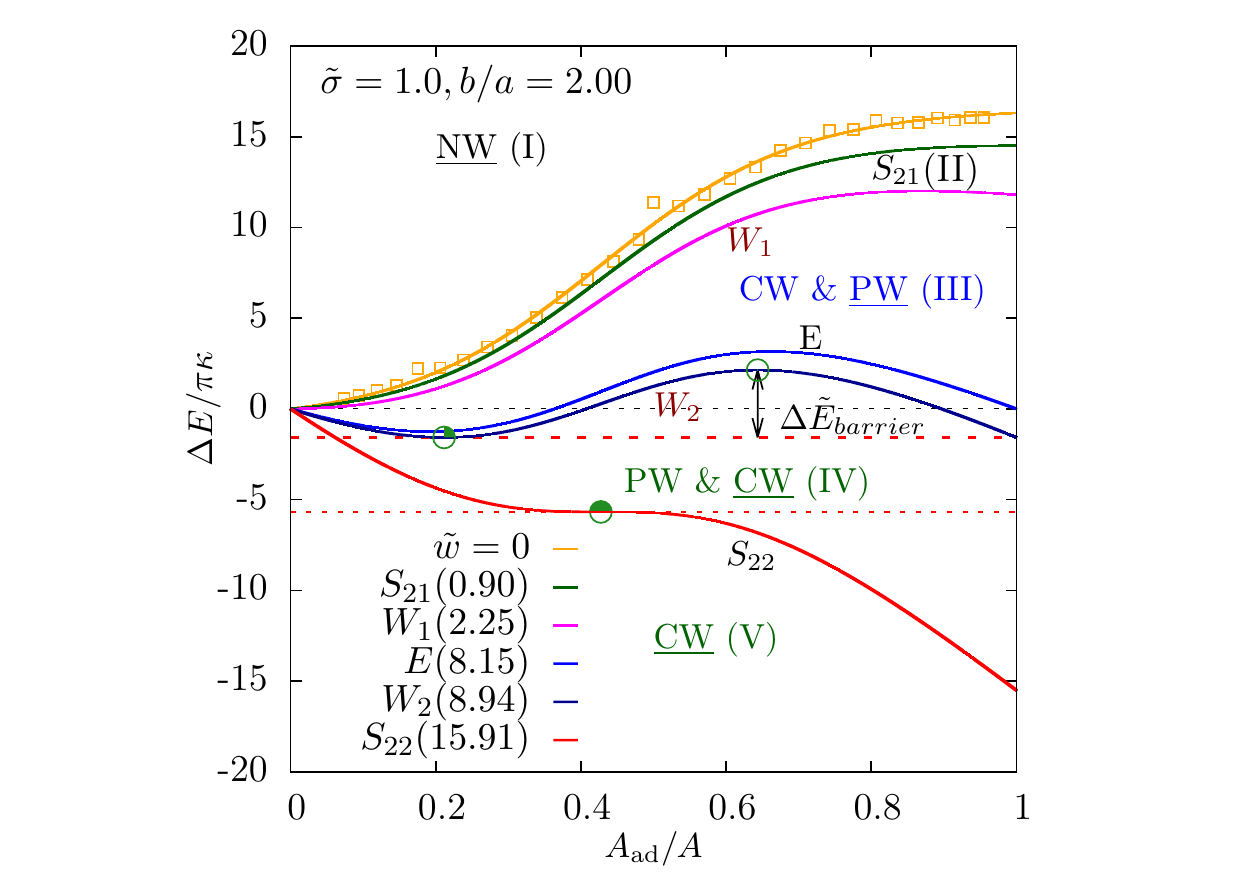}  \\[2ex]
\raisebox{1.4cm}{(b)} & \includegraphics[trim= 1.5cm 1.4cm 0.5cm 0cm, clip=true, totalheight=0.18 \columnwidth]{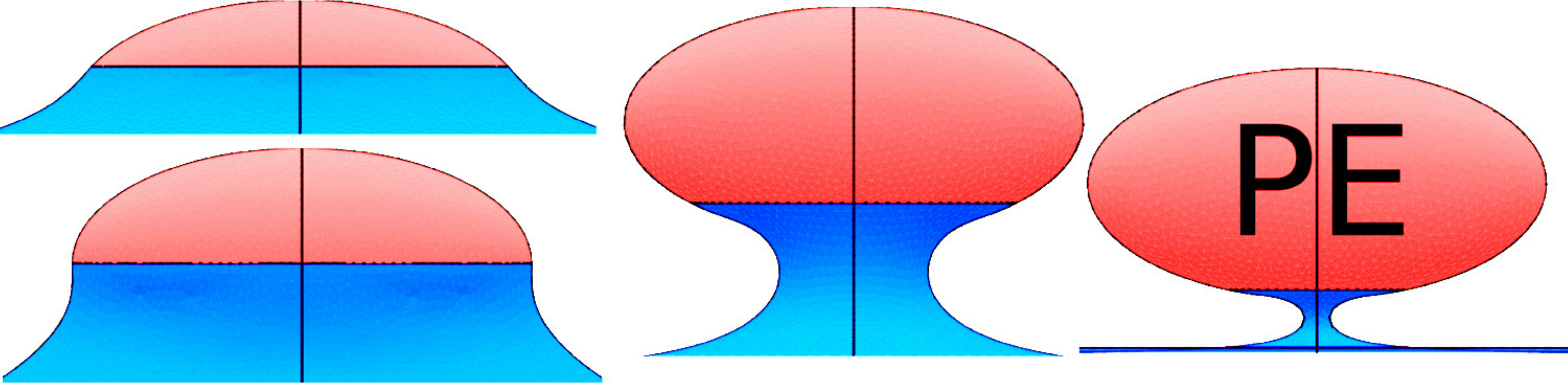} \\[2ex]
\raisebox{6.4cm}{(c)} & \includegraphics[trim= 1.8cm 0.1cm 2.2cm 0cm, clip=true, totalheight=0.8 \columnwidth]{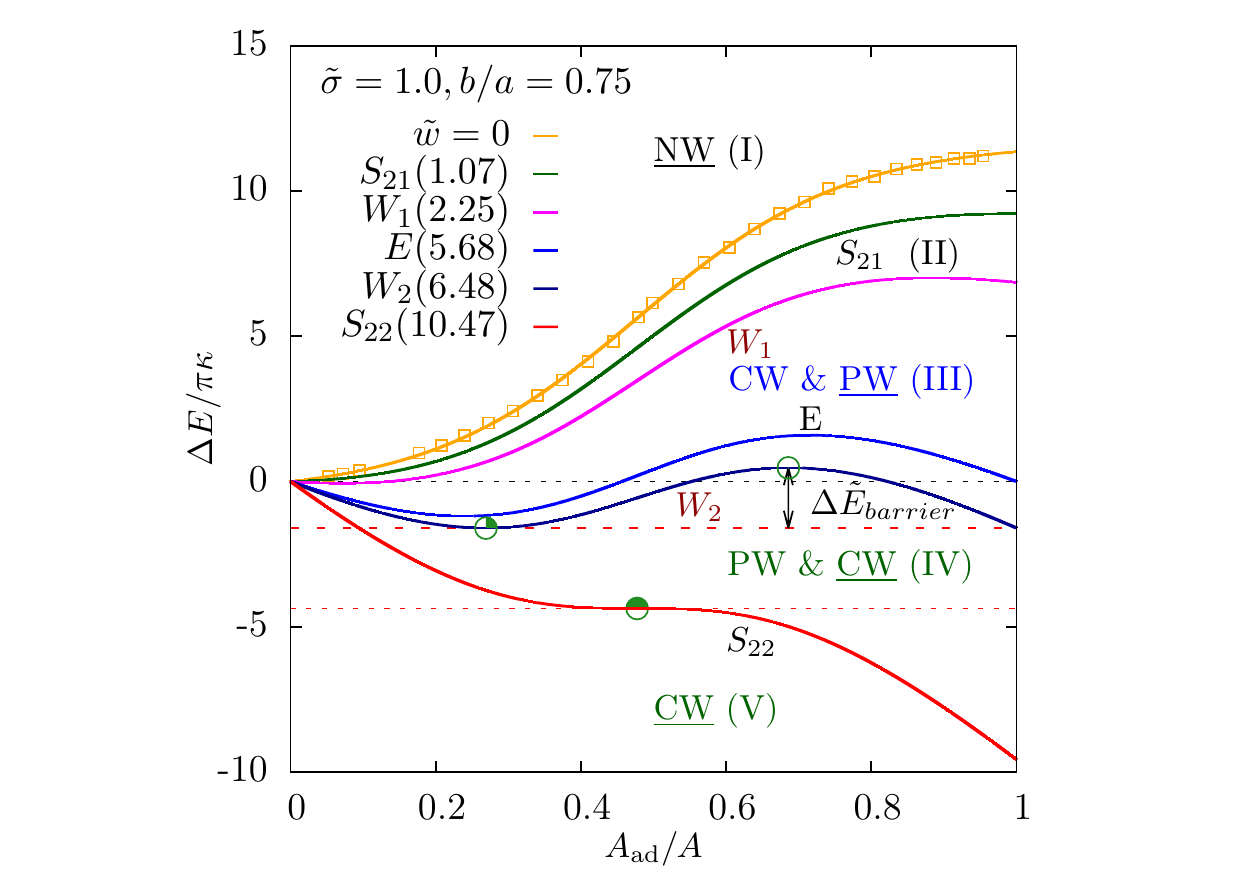}  \\[2ex]
\raisebox{1.4cm}{(d)} & \includegraphics[trim= 2cm 2cm 1.2cm 0cm, clip=true, totalheight=0.18 \columnwidth]{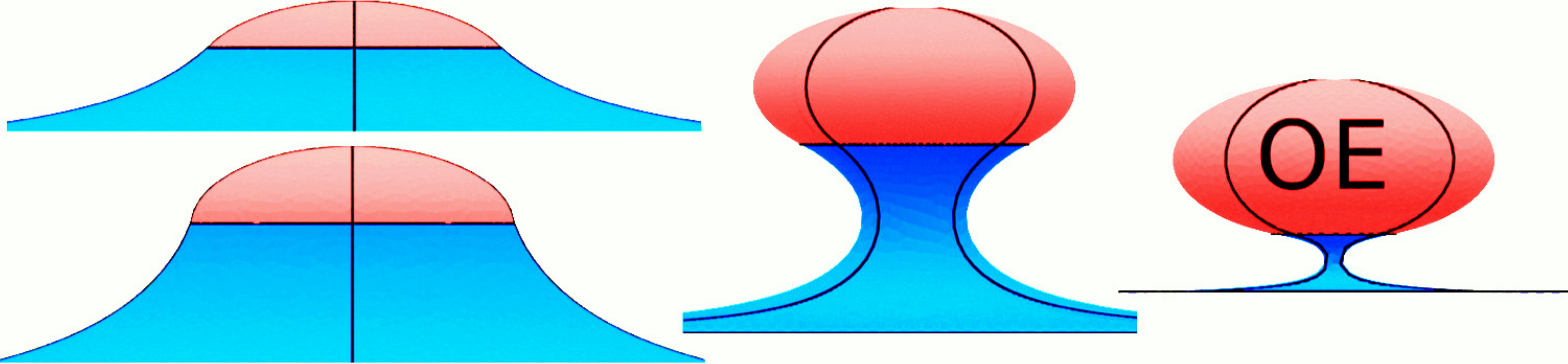}
\end{tabular}
\caption{Wrapping energies for (a) a prolate ellipsoid (PE) with aspect ratio $2$ and (c) an oblate ellipsoid (OE) with aspect ratio $0.75$, analogous to Fig.~\ref{WEDS} for a sphere. (b) and (d) Sketches for ellipsoids with wrapping fractions 0.30, 0.49, 0.84 and 0.95.  }
\label{WEDPEOE}
\end{figure}

The numerical data for the deformation energy without any adhesion  in Figs.~\ref{WEDS}~(a), \ref{WEDPEOE}~(a), and \ref{WEDPEOE}~(c) is fit by the function given in Eq.~(\ref{fiteq}). For increased adhesion strength, the onset of wrapping occurs for the adhesion strength $w_1$ for that the bending-energy cost equals the adhesion-energy gain at the contact point, see appendix \ref{A1}. For all adhesion strengths $w < w_1$ the stable state for the particle is the unwrapped state. Increasing the adhesion strength further, we find the envelopment transition from the partially-wrapped to the completely-wrapped state at adhesion strength $w_2$. For adhesion strengths $w_1 < w < w_2$, a global minimum of the wrapping energy is found for a small wrapping fraction, $0 < A_{\rm ad}/A \ll 1$. For adhesion strengths $w>w_2$, a stable completely-wrapped state is found. The line E, where the energy of the unwrapped state equals the energy of the completely-wrapped state is a good estimate for the binodal ${W_\mathrm{2}}$.

Whereas there is a continuous transition from the unbound to the bound state, the envelopment transition between the partially-wrapped and completely-wrapped state is characterized by an energy barrier, $\Delta \tilde {E}_{\rm barrier}$. For $w_1 < w < w_2$, in addition to the stable partially-wrapped state at small wrapping fraction, a metastable completely-wrapped state is found; similarly at higher adhesion strengths, in addition to the completely wrapped state a metastable partially-wrapped state is found. Indicated by the spinodal ${S_\mathrm{22}}$ that is associated with ${W_\mathrm{2}}$, the energy barrier between the metastable partially-wrapped state and the completely-wrapped state vanishes at an adhesion strength that is even larger than $w_2$. Starting from a completely-wrapped state and continuously decreasing the adhesion strength, below a threshold value $w_1$ a spontaneous transition between the completely-wrapped state and the unwrapped state is observed, which is indicated by the second spinodal for ${W_\mathrm{2}}$, ${S_\mathrm{21}}$ \footnote{Our numerical calculations and the fit function cannot capture a partially-wrapped at very high wrapping fraction, see Ref.~\onlinecite{deserno04}. Therefore the adhesion strength for that ${S_\mathrm{21}}$ occurs is obtained from the condition that the slope of the energy is zero at complete wrapping.}. The system shows strong hysteresis, such that no partially wrapped state with small wrapping fraction is encountered during this unwrapping transition. The height of the energy barrier between the partially-wrapped and the completely-wrapped state, $\Delta \tilde {E}_{\rm{barrier}}$, can be characterized by barrier height for $w=w_2$, as indicated in Figs.~\ref{WEDS} and \ref{WEDPEOE}. High energy barriers for ellipsoids with increased aspect ratio lead to an increased stability of the partially-wrapped state, compare section \ref{sec3c}.

\begin{figure} [t]
\centering
 \includegraphics[trim= 1cm 0.1cm 1cm 0.1cm, clip=true, totalheight=0.83\columnwidth]{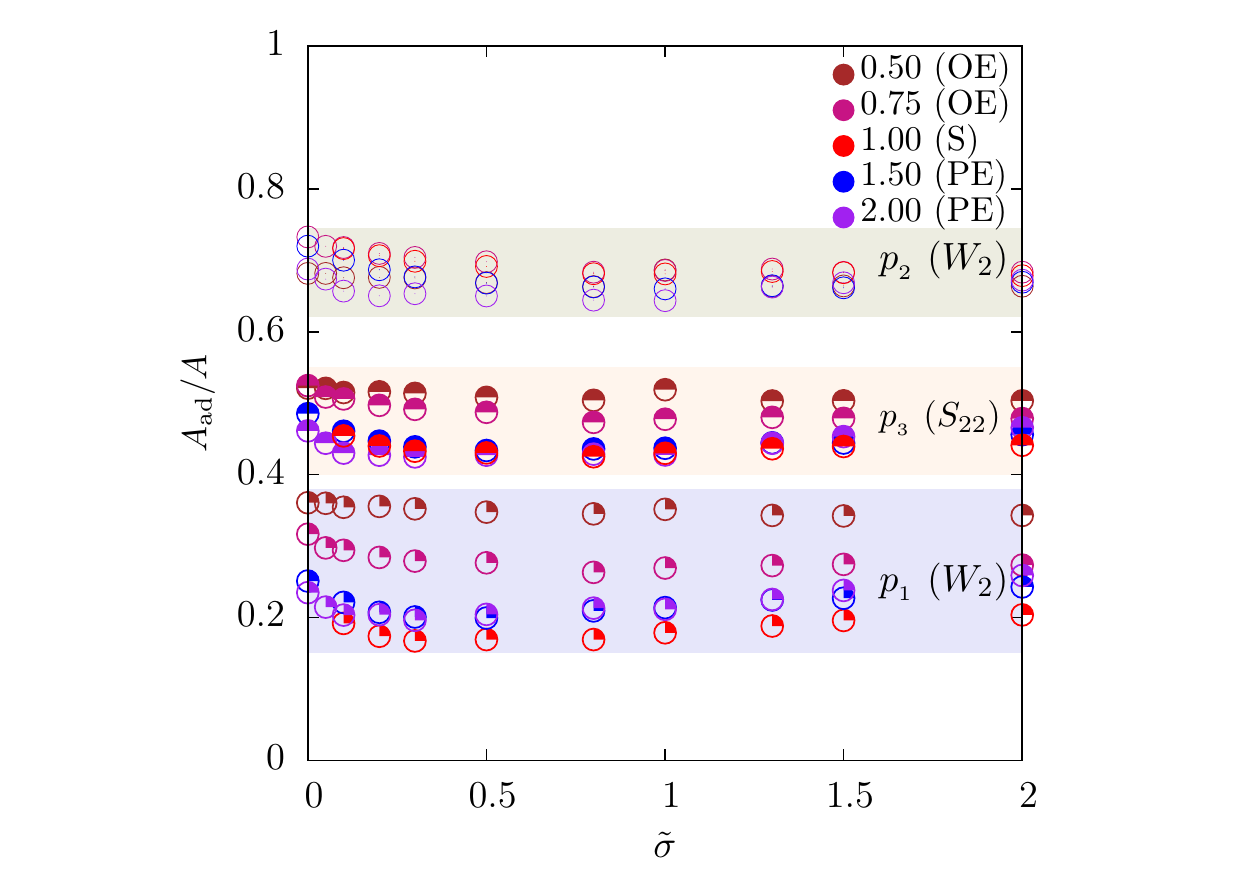}   
\caption{Wrapping fractions for special points marked on the wrapping-energy functions in Figs.~\ref{WEDS} and \ref{WEDPEOE} for several values of the reduced surface tension of the membrane and for several aspect ratios. At adhesion strength $w_2$, the partially-wrapped state is found at wrapping fraction $p_\mathrm{1}$ and the energy maximum at wrapping fraction $p_\mathrm{2}$; $p_\mathrm{3}$ is the wrapping fraction for that the energy barrier between partially-wrapped and completely-wrapped state vanishes.}
\label{PWF}
\end{figure}

Special wrapping fractions are marked by points in Figs.~\ref{WEDS} and \ref{WEDPEOE}, and are plotted  in Fig.~\ref{PWF} as function of the surface tension for several aspect ratios. The wrapping fraction $p_\mathrm{1}$ denotes the wrapping fraction for the stable partially-wrapped state, while $p_\mathrm{2}$ is the wrapping fraction for the energy maximum, both for adhesion strength $w_2$. The saddle point of the energy, when the barrier between the partially-wrapped and the completely-wrapped state vanishes on the spinodal ${S_\mathrm{22}}$, occurs at wrapping fraction $p_\mathrm{3}$. All wrapping fractions for special points depend only weakly on the surface tension, but decrease slightly with the increasing surface tension at small surface tensions.

When the partially-wrapped state has the same energy as the completely-wrapped state, the wrapping fractions for the partially-wrapped state are always well below $0.5$. However, a strong dependence of the wrapping fraction on the particle shape is observed: whereas for prolate ellipsoids and spheres $p_\mathrm{1} \approx 0.2$, for oblate ellipsoids it increases strongly with decreasing aspect ratio. The maxima of the energy barriers are found for wrapping fractions between $0.6$ and $0.7$. For adhesion strengths $w>w_2$ the partially-wrapped state becomes metastable; in this regime, the wrapping fraction at which the metastable partially-wrapped state appears increases with increasing adhesion strength until it reaches the critical value  $p_\mathrm{3}$, which is found at approximately half wrapping.

\subsection{Wrapping diagrams}
\label{sec3b}

\begin{figure} [t]
\centering
\includegraphics[trim= 1.5cm 0.1cm 1cm 0.1cm, clip=true,
totalheight=0.8\columnwidth]{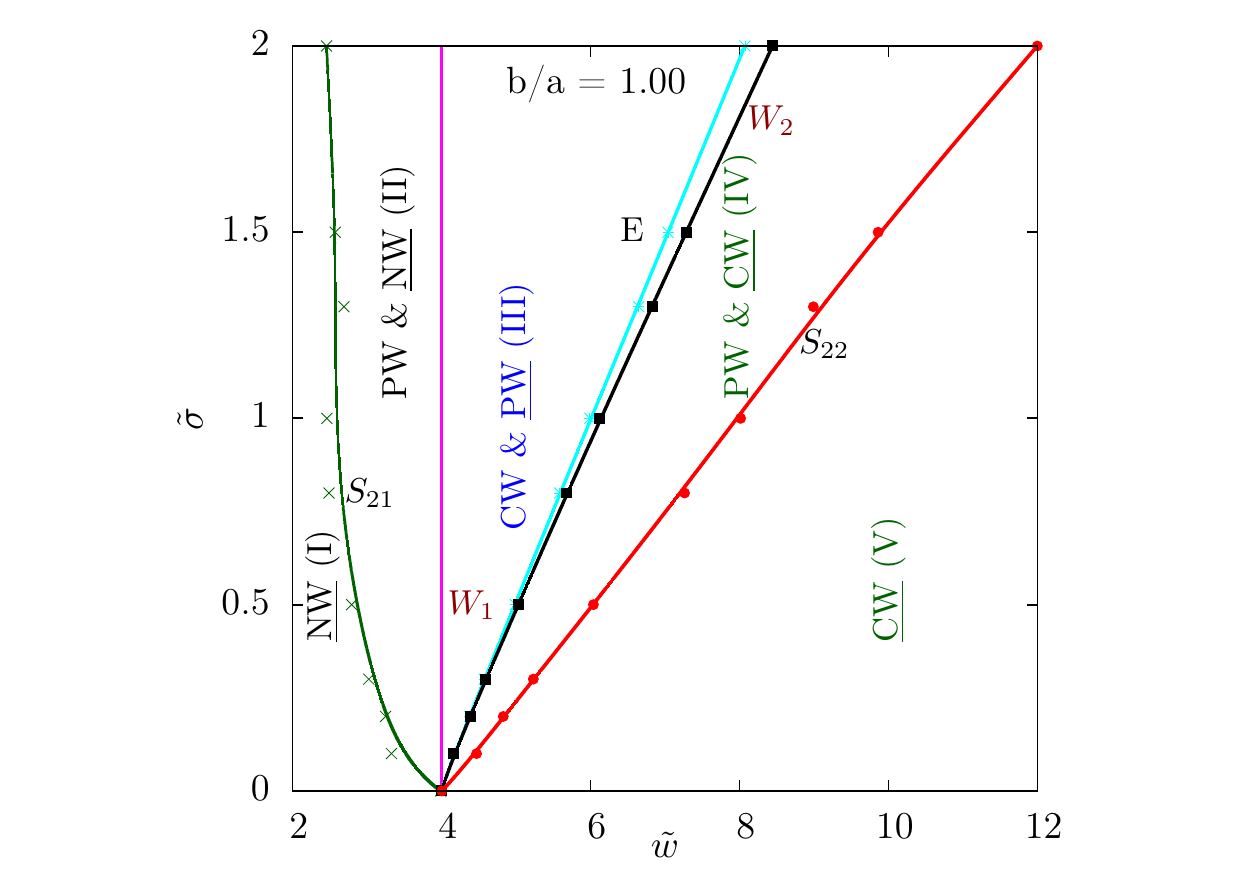}   
\caption{Wrapping states for spherical particles with reduced adhesion strength $\tilde w$ and reduced surface tension $\tilde \sigma$. Non-wrapped (NW), partially-wrapped (PW), and completely-wrapped (CW) states are found, stable and metastable states are indicated, stable states are underlined. The binodal ${W_\mathrm{1}}$ separates a stable non-wrapped and a stable partially-wrapped state, the binodal ${W_\mathrm{2}}$ separates a stable partially-wrapped and a stable completely-wrapped state. The two spinodals ${S_\mathrm{21}}$ and ${S_\mathrm{22}}$ belong to ${W_\mathrm{2}}$. For all states on $E$, the unwrapped state has the same energy as the completely-wrapped state.}
\label{SPD}
\end{figure}

\begin{figure} [t]
\centering
 (a)\includegraphics[trim= 1.5cm 0.1cm 1cm 0.2cm, clip=true,
totalheight=0.8\columnwidth]{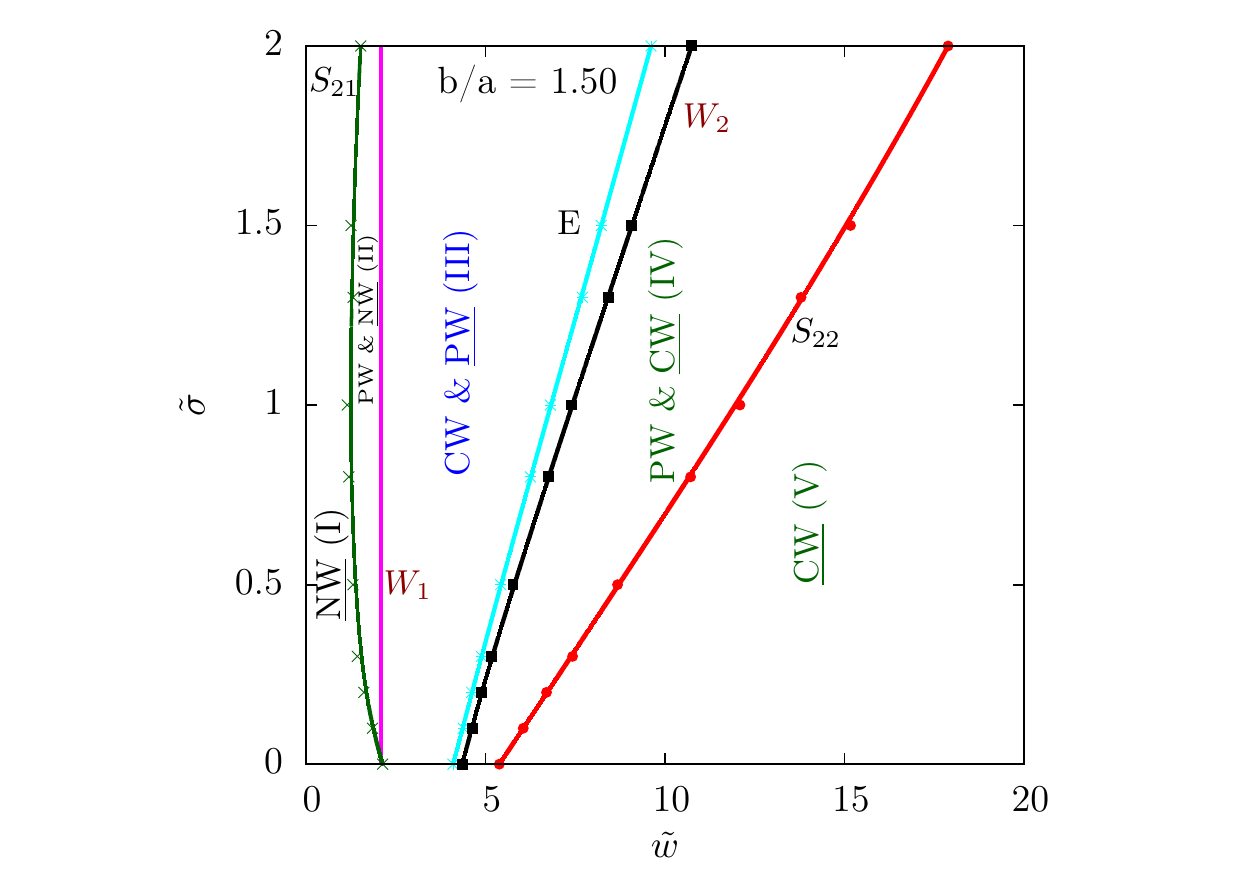}    \\
 (b)\includegraphics[trim= 1.5cm 0.1cm 1cm 0.2cm, clip=true,
totalheight=0.8\columnwidth]{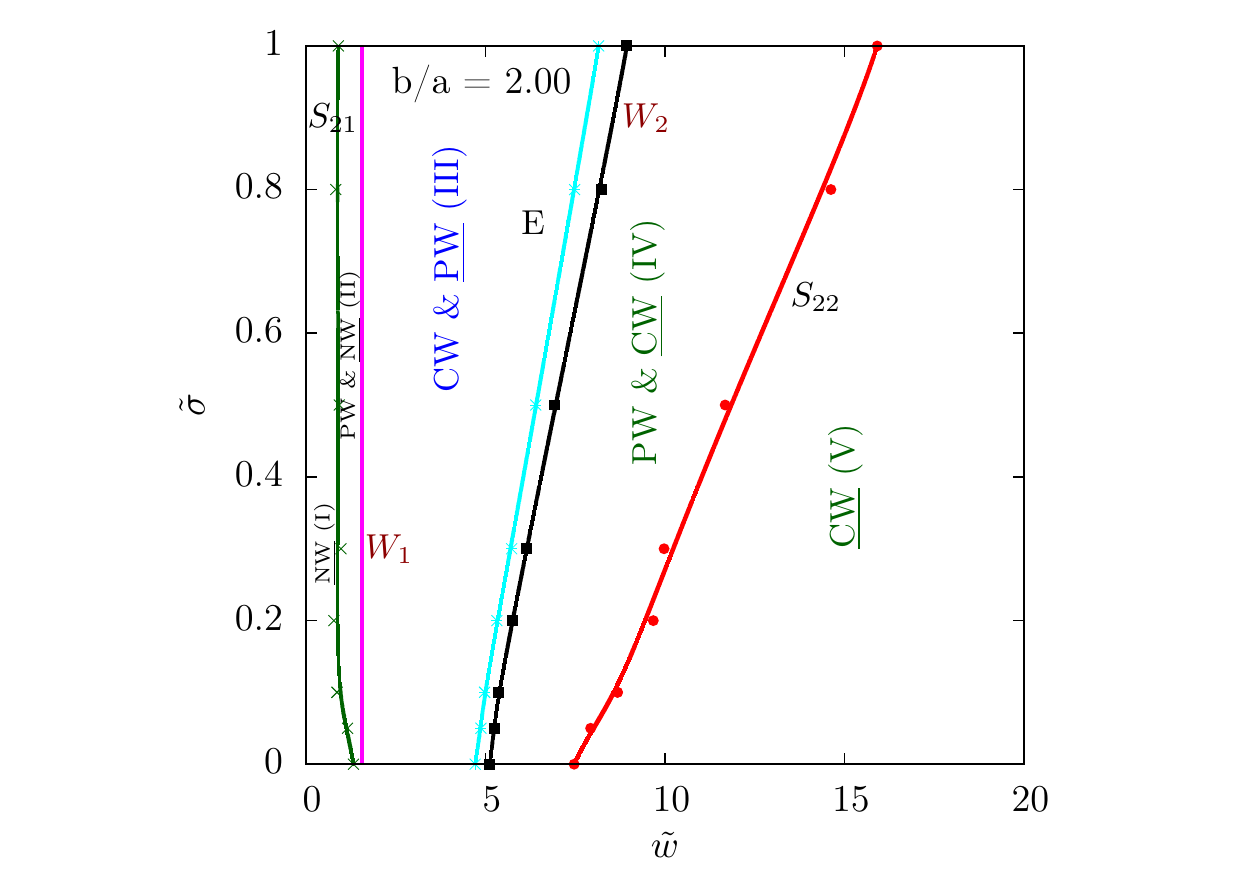} 
\caption{Wrapping states for prolate ellipsoidal particles with aspect ratios (a) $1.5$ and (b) $2$, plotted analogously to Fig.~\ref{SPD}.}
\label{PEPD}
\end{figure}

\begin{figure}[t]
\centering
 (a)\includegraphics[trim= 2cm 0.1cm 1cm 0.2cm, clip=true, totalheight=0.8\columnwidth]{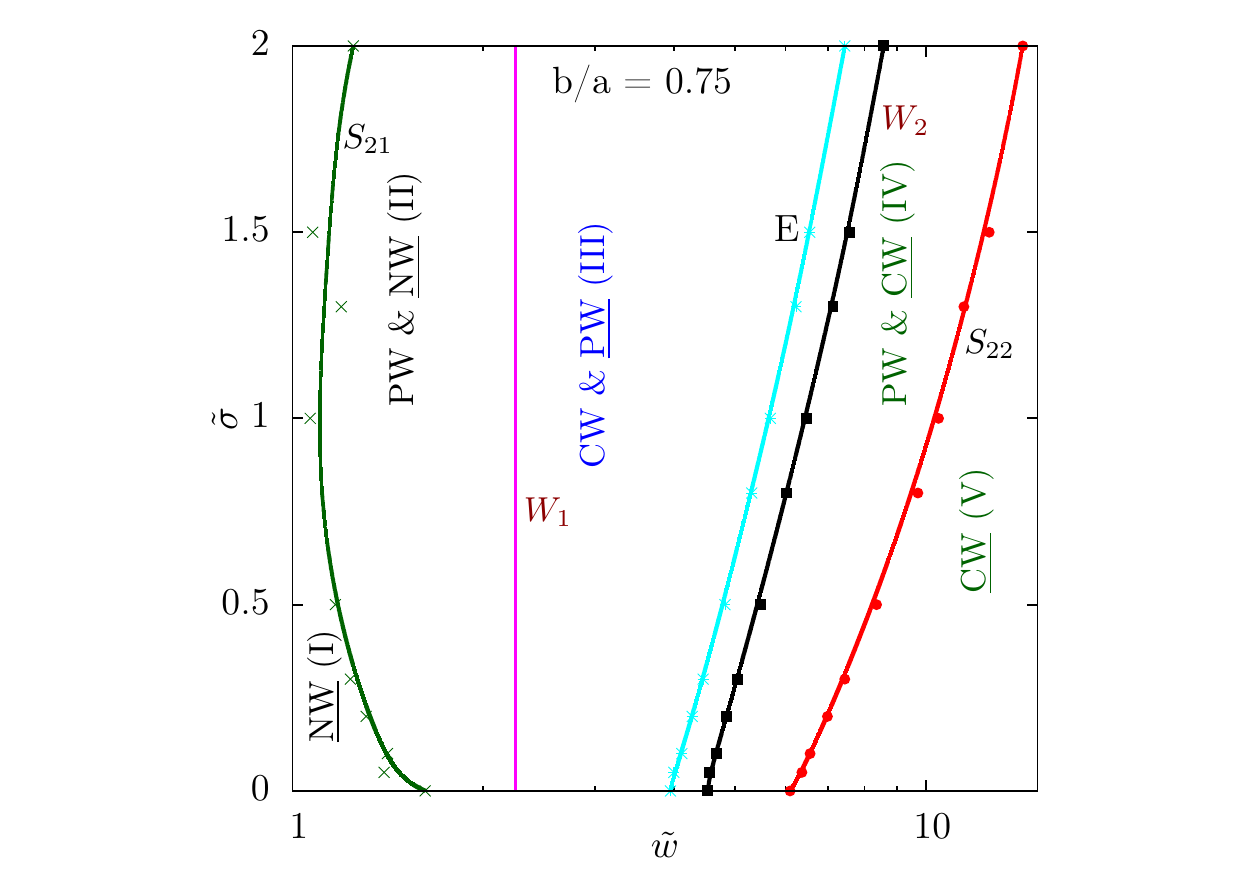}   \\
 (b)\includegraphics[trim= 2cm 0.1cm 1cm 0.2cm, clip=true, totalheight=0.8\columnwidth]{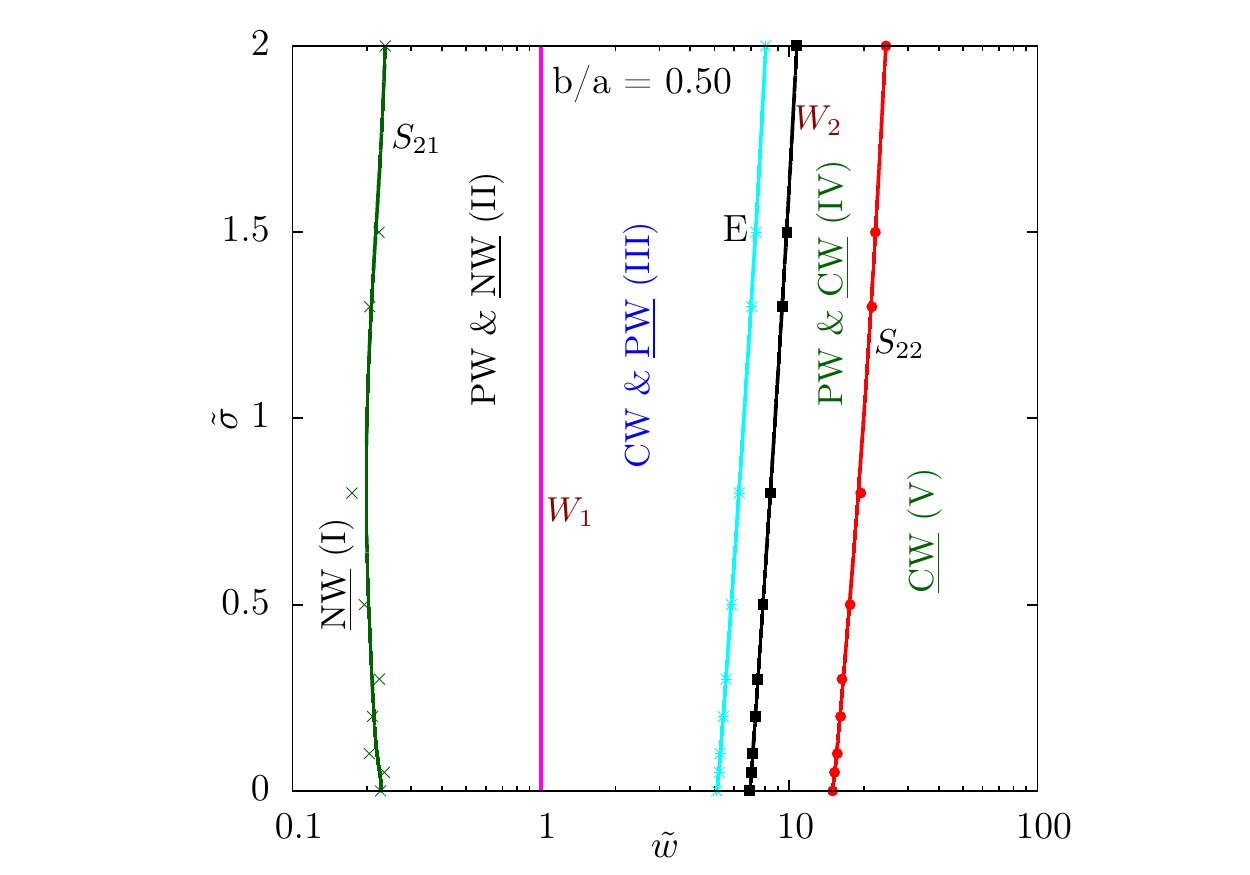} 
\caption{Wrapping states for oblate ellipsoidal particles with aspect ratios (a) $0.75$ and (b) $0.5$, plotted analogously to Fig.~\ref{SPD}.}
\label{OEPD}
\end{figure}

For wrapping a spherical particle by a tension-less membrane, the only relevant energies that determine the uptake process are bending and adhesion energy for the adhered part. At every point of the sphere the bending energy is constant and the free membrane around the particle forms a catenoid-like deformation, i.~e., a minimal surface with vanishing bending energy cost. Thus there exist only $2$ possible stable states, viz. a non-wrapped and a completely-wrapped state, that are separated by a continuous transition at $\tilde w = 4$. For a spherical particle and a finite membrane tension, also a stable partially-wrapped state is found, compare section \ref{sec3a}. For ellipsoidal particles this partially-wrapped state exists both for finite and for vanishing surface tension. 

In the wrapping diagrams for surface tension and adhesion strength in Figs.~\ref{SPD}, \ref{PEPD}, and \ref{OEPD}, five regimes can be identified with different combinations of stable and metastable unwrapped, partially wrapped, and completely wrapped states. For small adhesion strengths, a stable non-wrapped state is found. In between the spinodal for the spontaneous transition between the completely-wrapped and the non-wrapped state ${S_\mathrm{21}}$, and the binodal for the binding of the colloid to the membrane ${W_\mathrm{1}}$, in addition to the stable non-wrapped state a metastable completely-wrapped state with high wrapping fraction appears \footnote{It should be a partially-wrapped state at high wrapping fraction \cite{deserno04}, but this cannot be described by the function in Eq.~(\ref{fiteq}).}. The binding transition occurs at adhesion strength $w_{1,\rm PE}= \left[ 1 + (b/a)^{-2} \right]^2$ and $w_{1,\rm OE} = 4(b/a)^2$ for prolate and oblate ellipsoids respectively and is independent of the surface tension, see appendix \ref{A1} and Refs.~\onlinecite{deserno04,seifert1990, Lipowsky_Dobereiner_1998}.  Beyond the threshold adhesion strength for binding, a partially-wrapped state coexists with the metastable completely-wrapped state. For adhesion strengths beyond those for the binodal ${W_\mathrm{2}}$, the completely-wrapped state becomes stable and the partially-wrapped state becomes metastable. Finally, beyond the spinodal ${S_\mathrm{22}}$, the energy barrier between the completely-wrapped and the partially-wrapped state vanishes.

For spherical particles, all phase boundaries meet at a triple point for zero surface tension, see Fig.~\ref{SPD}. However, as shown in Figs.~\ref{PEPD} and \ref{OEPD} for ellipsoids, even for zero surface tension ${W_\mathrm{1}}$ is shifted to smaller values of the adhesion strength compared with spherical particles, while the binodal ${W_\mathrm{2}}$ is shifted to higher values of the adhesion strength. For comparable aspect ratios, this effect is stronger for oblate ellipsoids than for prolate ellipsoids.
This range of adhesion strengths, for which stable partially-wrapped states are found, increases both with the aspect ratio of the particle and the surface tension of the membrane. For increased aspect ratio it is easier to attach particles to the membrane, but at the same time it is more difficult to achieve completely-wrapped state. Binding occurs already for smaller adhesion strengths because of the reduced particle curvature at the point of first contact, whereas complete wrapping requires higher energies because of the increased curvature of the particle at the tips. For high aspect ratios and rather 'flat' particles, the dependence of the binodal ${W_\mathrm{2}}$ and both associated spinodals on the surface tension decreases.

\subsection{Energy barrier}
\label{sec3c}

\begin{figure} [t]
\centering
\includegraphics[trim= 1cm 0.1cm 1cm 0.2cm, clip=true, totalheight=0.8\columnwidth]{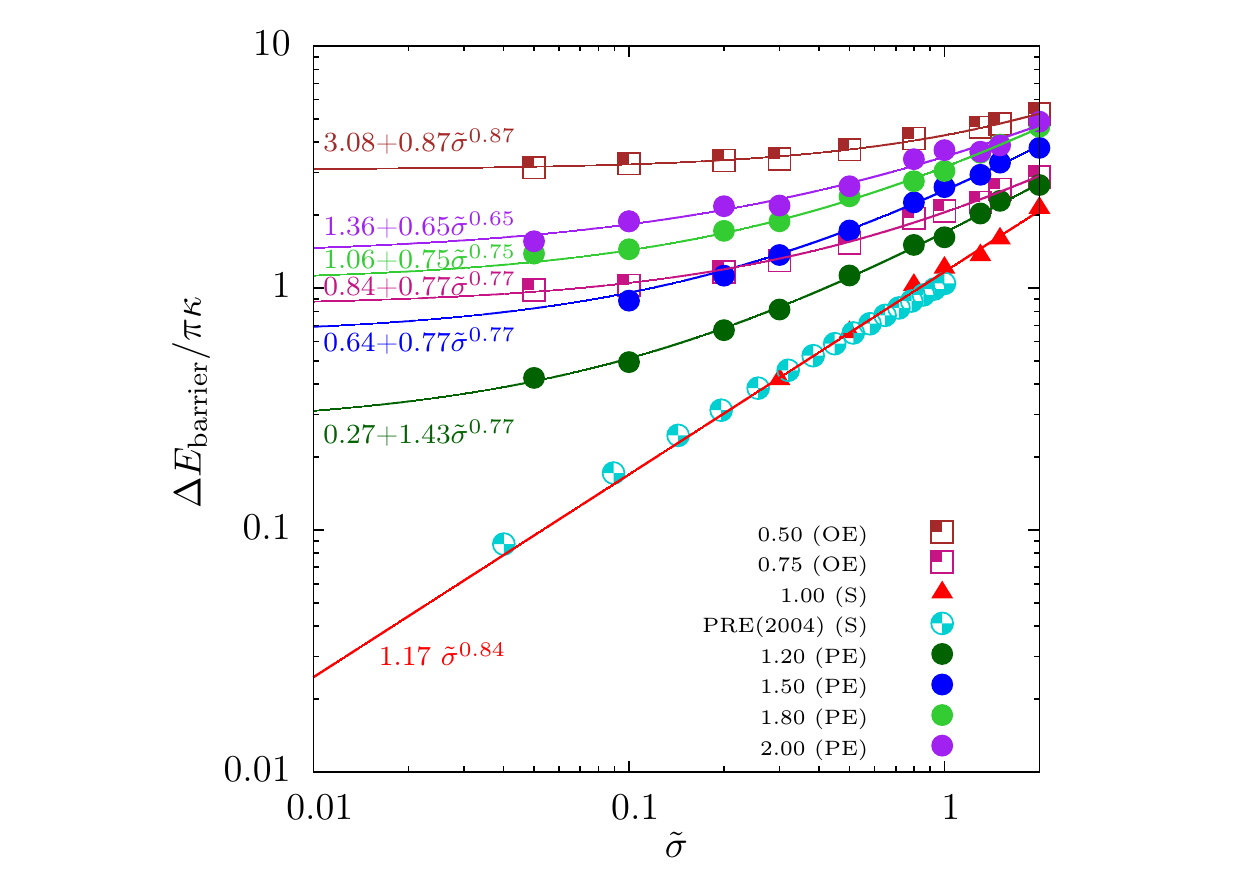}  
\caption{Energy barriers between partially-wrapped and completely wrapped state for the adhesion strength $w_2$ where partially and completely-wrapped state have equal energies. The barrier energies are plotted for several aspect ratios as function of the reduced surface tension $\tilde \sigma$ and fit by a function $\Delta \tilde E_{\rm barrier} = \tilde E _0 + \tilde E_1 \tilde{\sigma}^\nu$. The values for the sphere are compared with those from Ref.~\onlinecite{deserno04}.}
\label{EBAR}
\end{figure}

\begin{figure} [t]
\centering
\includegraphics[trim= 1.5cm 0.15cm 1.2cm 0.3cm, clip=true, totalheight=0.80\columnwidth]{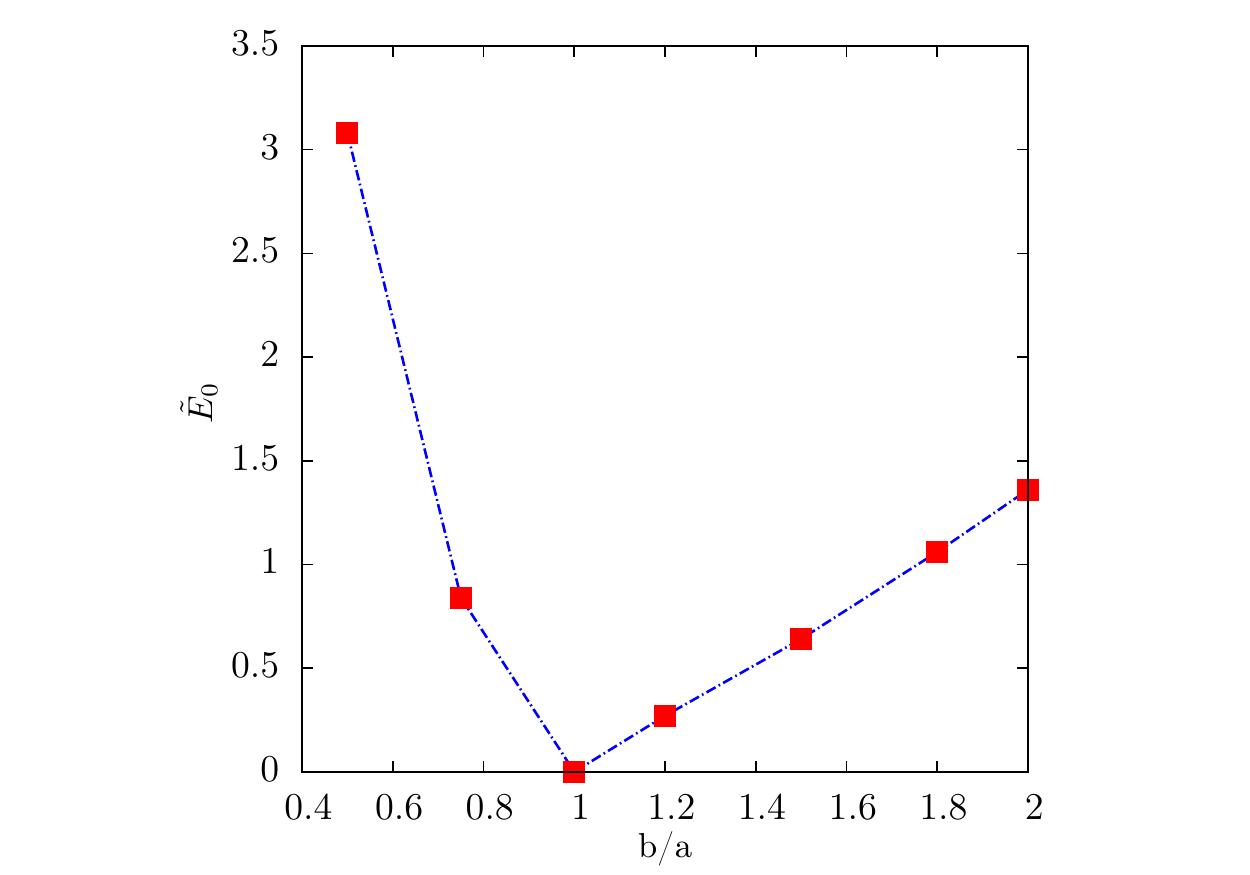}  
\caption{Energy barrier $\tilde E _0$  as a function of the aspect ratio b/a for vanishing surface tension. Oblate particles have b/a $<$ 1 and prolate particles have b/a $>$ 1. }
\label{E0AR}
\end{figure}

In Figs.~\ref{WEDS} and \ref{WEDPEOE}, the wrapping energy for adhesion strength $w_2$ shows a partially-wrapped state at a wrapping fraction $p_{1}$ that is separated by an energy barrier $\Delta \tilde {E}_\mathrm{barrier}$ from the completely-wrapped state. For $w>w_2$, the height of the barrier between a metastable partially-wrapped state and the completely-wrapped state determines the dynamics for the wrapping process. The activation energy that is needed for complete wrapping can be provided either by thermal kicks or by active forces from motor proteins. A comparison of the height of the energy barrier shown in Fig.~\ref{EBAR} with the thermal energy $k_{\mathrm{B}}T$ therefore allows to estimate a characteristic time for the transition \cite{Lipowsky92}.

The energy barriers vary with the surface tension of the membrane and with particle shape. While for spherical particles the energy barrier vanishes for the tension-free case and can be fit by a power law \cite{deserno04}, it levels off to finite values both for prolate and oblate ellipsoids. The energy barriers for ellipsoidal particles thus can be orders of magnitude higher than those for spherical particles. We fit the energy barriers as function of the wrapping fraction by $\Delta \tilde{E}_\textrm{barrier}= \tilde{E}_{0} + \tilde{E}_{1} \tilde{\sigma}^{\nu}$. While the exponent $\nu$ is similar for all particles, the prefactors $\tilde{E}_{0}$ and $\tilde{E}_{1}$ strongly depend on the particle shape. 

While in section \ref{sec3b} an extended parameter range for a partially-wrapped regime has been found for ellipsoidal particles compared with spherical particles, the energy barrier characterizes the stability of metastable partially-wrapped states. This can practically imply that partially-wrapped ellipsoidal particles are found in experiments even for adhesion strengths much higher than $w_2$, for that wrapping diagrams already predict a completely-wrapped particle. In Fig.~\ref {E0AR}, the energy barriers for vanishing surface tension are plotted as function of the aspect ratio of the particle. Only for $\tilde E _0 < k_{\mathrm{B}}T/(\pi\kappa)$ the barrier height is comparable to thermal energy and complete wrapping is expected to occur at  $w_2$. $\tilde {E}_{0}$ is of the order of $1$ already for prolate ellipsoids with aspect ratio $1.5$ thus corresponds to an energy barrier of about $\pi\kappa$. This implies that  for particles whose shape considerably deviates from a sphere, adhesion strengths that are considerably higher than $w_2$ are required to wrap the particle and the partially-wrapped regime might extend almost up to the spinodal $S_{22}$.

\subsection{Role of shape anisotropy}
\label{sec3d}

\begin{figure} [t]
\centering
 \includegraphics[trim= 0cm 0cm 0cm 0cm, clip=true,totalheight=0.8\columnwidth]{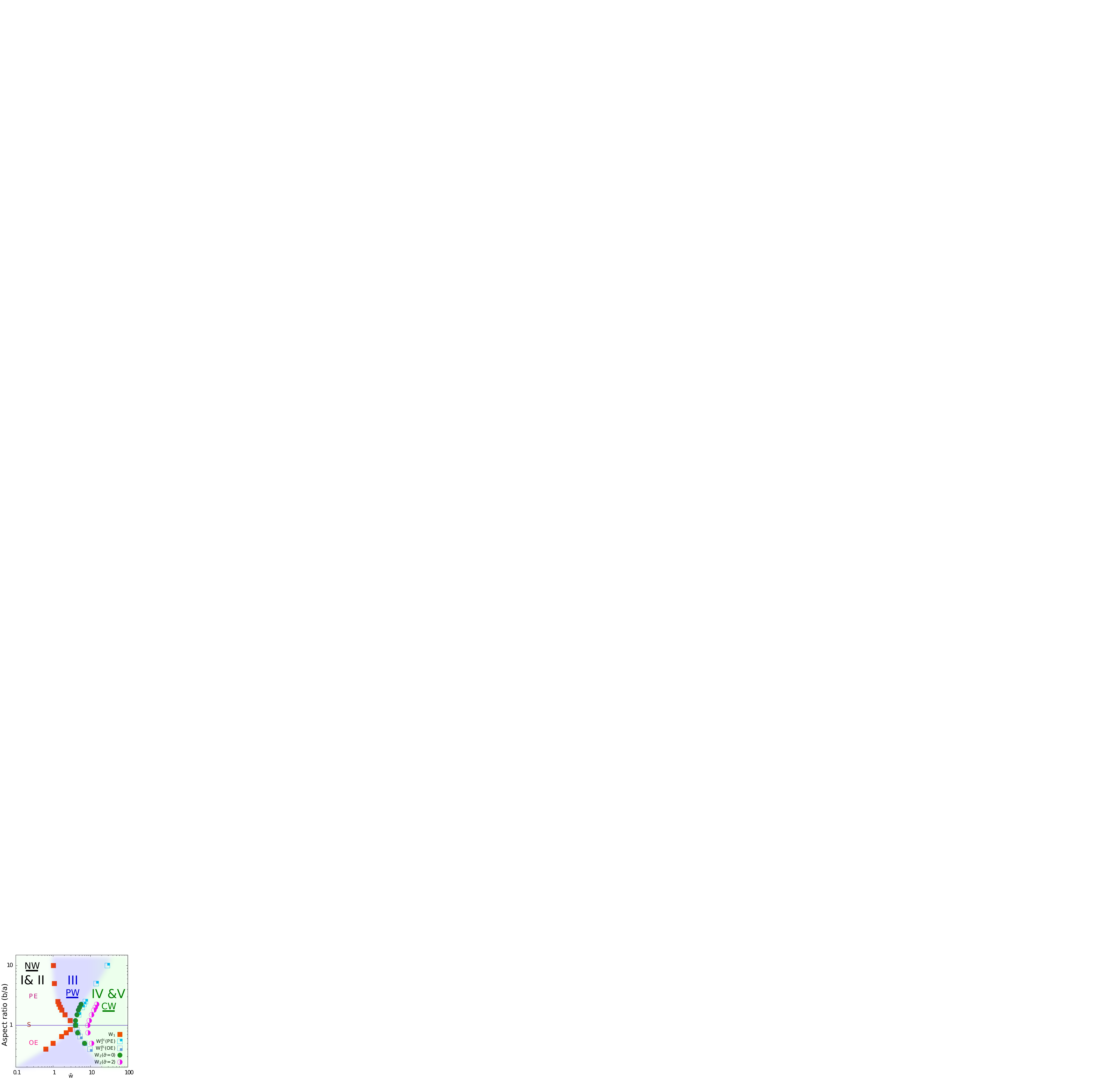}  
 \caption{Stable wrapping states for different aspect ratios $b/a$, reduced adhesion strengths $\tilde w$, and for $\tilde \sigma = 0$ and for $\tilde \sigma = 2$. Prolate  ellipsoids (PE) have aspect ratios $b/a>1$, while for oblate ellipsoids (OE) have aspect ratios $b/a<1$. Stable wrapping states are the non-wrapped state (NW), the partially-wrapped state (PW), and the completely-wrapped state (CW); the roman numbers correspond to those used in Figs.~\ref{SPD}, \ref{PEPD}, and \ref{OEPD}. For blue open squares labeled with index $\rm dc$, the deformed catenoid approximation has been used.}
\label{ARPD}
\end{figure}

Stable non-wrapped, partially-wrapped and completely-wrapped regimes can be identified in Fig.~\ref{ARPD}, where the state of the system depends on membrane surface tension, particle aspect ratio, and adhesion strength. For spherical particles, the transition from the non-wrapped to the completely-wrapped state occurs at the triple point for $\tilde w=4$. For ellipsoids, the transition from the non-wrapped to the partially-wrapped state is found already for $w_1<4$, where $w_{1, \rm PE} \rightarrow 1$ for prolate ellipsoids with high aspect ratio and $w_{1,\rm OE} \rightarrow 0$ for disc-like oblate ellipsoids, compare Eqs.~(\ref{eq:W1PE}) and (\ref{eq:W1OE}). The limiting cases for $w_2$ correspond to an infinite cylinder \cite{Weikl03, Mkrtchyan_Chen10} and a flat disc \cite{nowak08,Gao11} respectively. For the envelopment transition, we find an increased adhesion strength $\tilde w_2$ for the prolate and oblate ellipsoids with increasing asphericity; an additional shift towards higher adhesion strength is observed for finite surface tension.

Besides the numerical minimization technique to calculate $w_2$, we have used a deformed-catenoid approximation to estimate the wrapping energy that allows to estimate $w_2$ even for high aspect ratios such as $10$. For a tensionless membrane, we numerically calculate the deformation energy for a wrapped ellipsoid by deforming a sphere with a catenoidal membrane patch, without any actual minimization. This method works well, because without surface tension the contribution of the deformation of the membrane patch around the particle is small. For oblate particles and bending-only, the membrane patch around the particle will assume a catenoid shape with vanishing bending energy and therefore does not have to be calculated.

As discussed in section~\ref{sec3c}, for $ \tilde {w}>$${W_\mathrm{2}}$ the stable state is the completely-wrapped state (\underline {CW}), but still the particles may be stuck in metastable partially-wrapped states unless there are sufficiently large fluctuations such that the energy barrier can be crossed, compare Fig.~\ref{EBAR}. Thus, for a given adhesion strength that is smaller than the one
for the spinodal for spontaneous wrapping,
we predict a considerably smaller amount of completely-wrapped particles the stronger the asphericity is.

\section{Summary \& Conclusions}

Complete wrapping a spherical particle using a tensionless free membrane occurs at reduced adhesion strength $\tilde w = 4$, directly from the unwrapped to the wrapped state without any energy barrier. For a membrane with surface tension the envelopment transition is shifted to higher adhesion strengths, while the binding transition remains at $\tilde w = 4$. The new partially-wrapped state is separated from the completely-wrapped state by an energy barrier. We find that a partially-wrapped state also exists for ellipsoidal particles, in a wider region of the phase diagram and with a higher energy barrier to the fully-wrapped state than for spherical particles. Therefore, the spherical shape facilitates complete wrapping for single particles; a biological example is the uptake and budding of almost spherical viruses \cite{Sun_Wirtz_2006,Zhang_Nguyen08}. 

In addition to a shift due to a membrane surface tension, for ellipsoidal particles the envelopment transition additionally shifts to higher adhesion strengths with increasing non-sphericity. The binding transition shifts to smaller adhesion strengths: for very long prolate ellipsoids $w_1 \rightarrow 1$ and for very flat oblate ellipsoids to $w_1 \rightarrow 0$. The partially-wrapped state is additionally stabilized by a higher energy barrier to the completely-wrapped state. Therefore, attachment of ellipsoidal particles to a membrane and partial wrapping is facilitated compared with spheres, while complete wrapping is hindered; typical wrapping fractions are $20-40\%$. Elongated viruses are found to form patterns on the membrane, see Ref.~\onlinecite{Kubo_2009}. However, similar to curved inclusions the particles may bud cooperatively \cite{auth09,reynwar07}.

For typical lipid bilayer bending rigidities of $\kappa = 20 \, k_{\mathrm{B}}T$, the reduced energies $\tilde E \approx 1$ correspond to typical energies $E \approx 60$ $k_{\mathrm{B}}T$. Energy barriers at adhesion strengths where the energy of the partially-wrapped state equals the energy of the completely wrapped state can thus be of the order of $100 \, k_{\mathrm{B}}T$.
Often the length scale for the particle in soft matter and biological system is of the order of hundred nanometer, therefore $\tilde \sigma = (a^2/\kappa) \sigma = 500$ $\sigma  \, \rm{nm^2}/$$k_{\mathrm{B}}T $; a biologically relevant surface tension for the cell membrane, $\sigma = 0.003$ $\rm dyn/cm$ \cite{Morris_Homann_2001} corresponds to $\tilde \sigma \approx 1$ for a particle size $a = 100 \, \rm nm$, to  $\tilde \sigma \approx 0.25$ for a particle size $a = 50 \, \rm nm$, and to $\tilde \sigma \approx 4$ for a particle size $a = 200 \, \rm nm$.
At a reduced tension $\tilde \sigma = 0.25$, the energy barrier to the completely wrapped state is an order of magnitude higher for prolate ellipsoids of aspect ratio $2$ and oblate ellipsoids of aspect ratio $0.5$ than for a spherical particle. 

A typical adhesion strength can be estimated based on the binding strength of the HIV virus, $w \approx 0.1 \,k_{\mathrm{B}}T/\rm{nm}^2$ \cite{Sun_Wirtz_2006}. The reduced adhesion strength is thus  $\tilde w = 2 (a^2/\kappa) w = 100$ for a particle with size  $a = 100 \, \rm nm$, $\tilde w \approx 25$ for $a = 50 \, \rm nm$, and to $\tilde w \approx 400$ for $a = 200 \, \rm nm$. All adhesion strengths are well in the region where wrapping occurs. However smaller adhesion strengths may occur for other viruses or smaller receptor densities in the membrane.
An increased reduced tension hinders budding, but because $\tilde \sigma$ and $\tilde w$ both scale with the squared size of the particle, according to our calculations and previous wrapping calculations larger nano-particles are more likely to be wrapped completely. A lower limit for uptake is $\approx 20 \, \rm nm$, when the adhesion energy balances the bending energy \cite{Canton_Battaglia_2012,Zhang_Suresh_2009,Gao05, chaudhuri11}. However, there is also an upper limit given either by the length scale $\sqrt{\kappa/\sigma}$ where the wrapping becomes surface-tension dominated \cite{deserno04} or by receptor availability \cite{Zhang_Suresh_2009,Gao05, chaudhuri11}.  

Our theoretical calculations predict an enhanced stability of partially-wrapped states for ellipsoidal particles. For cells, a lower uptake for ellipsoidal particles has been found experimentally \cite{Chithrani06, Florez_landfester12,Champion_Mitragotri06}, partially adhered ellipsoidal particles are also discussed in particular in Ref.~\onlinecite{Florez_landfester12}, similarly disk-shaped particles show an increased adhesion to membrane and lower uptake \cite {Adriani_Decuzzi_2012,Zhang_Yan_2012}. Receptor-mediated wrapping of ellipsoidal particles has been studied in Ref.~\onlinecite{Decuzzi_Ferrari_2008} with similar findings as in our work, but without surface tension of the membrane. The partially wrapped state can be of advantageous both from an application point of view as well as from a biological point of view: for example, elongated particles can be used as markers for imaging that stay in the cell membrane \cite{Xu_2011,Pissuwan_Cortie_2008} and Ebola and Marburg viruses are not easily taken up by macrophages and may thus have a high virulence \cite{Canton_Battaglia_2012}. 

\section*{ACKNOWLEDGMENTS}
We thank K. Brakke (Selinsgrove, PA) for his advice on Surface Evolver and for helpful discussions on numerical techniques, and R. Korenstein (Tel Aviv) for
stimulating discussions on nano-particles at membranes.  Support from the EU FP7 NMP collaborative project PreNanoTox (309666) 
is gratefully  acknowledged. SD acknowledges support by the International Helmholtz Research School of Biophysics and Soft Matter (IHRS BioSoft).

\appendix
\section{Triangulated membranes}
\label{A2}

Triangulated membranes are a powerful tool to study membranes and interfaces \cite{kantor1989,Brakke92,gompper92,Wintz_Dobereiner_Seifert_1996,gompper1997,sunilkumar01,Kohyama_Kroll_Gompper03,fosnaric09}. Using Surface Evolver \cite{Brakke92}, the membrane shape can be minimized with different schemes and the triangulation can be refined at any stage. In this appendix, we present the discretization that has been used to calculate the energy in Eq.~(\ref{maineq}). Surface tension and adhesion energies are proportional to the area of the membrane, they can be calculated basically as sum over all triangle areas. The bending energy is calculated, using the Surface Evolver method ``star\_perp\_sq\_mean\_curvature'', which assumes every vertex  has incident triangles forming a star network around it, such that $a_{v}$ is the average area associated locally with the vertex. The force acting on a vertex when the area changes thereby causing the vertex to move is given by the gradient of the  area $(\boldnabla a_{v})$ associated with the vertex. For a smooth surface the gradient of the volume $(\boldnabla \mathcal{V}_{v})$ is equal to the magnitude of the area. But for a triangulated patch, the facets around the vertex tilt and thus the area is greater than the magnitude of the gradient of the volume. For a membrane patch, with a central vertex  $\bold v$ and neighboring vertices $\bold {v}^1$, $\bold {v}^2$, $\bold {v}^3$, ... $\bold{ v}^n$ a volume is given by   
 \begin{equation}
  {  \mathcal {V}_{v}}  = \frac{1}{6} \bold{v} \cdot \left[ {\bold {v}^1 \times \bold {v}^2 + \bold {v}^2 \times \bold {v}^3 +... \bold {v}^n \times \bold {v}^1}  \right] \, , 
\end{equation}
and the gradient of the volume is therefore
\begin{equation}
   \boldnabla \mathcal{V}_{v}  = \frac{1}{6}\left[ {\bold {v}^1 \times \bold {v}^2 + \bold {v}^2 \times \bold {v}^3 +... \bold {v}^n \times \bold {v}^1}  \right] \, . 
\end{equation}
The local mean curvature  ($h_v$) at the vertex is  
\begin{equation}
  {h}_{v}= \frac{1}{2}\frac{\boldnabla a_{v} \cdot \boldnabla \mathcal{V}_{v}}{ \boldnabla \mathcal{V}_{v} \cdot  \boldnabla \mathcal{V}_{v}} \, .
\end{equation}
Thus the discretized form of  total squared mean curvature integral  is given by,
\begin{equation}
\mathcal{E}_{\rm bend} = 2 \kappa \sum_{v=1}^{n} a_v h_v^2 \, .
\end{equation}

Once assembled, the surface can be minimized using different schemes, default being moving towards the direction of steepest descent of the energy linearly, while the mesh may be refined or smoothed at any stage. One may employ other minimization schemes like Hessian approach which calculates the energy of the surface configuration  for a small perturbation and then uses the Hessian, a   square matrix formed of the second derivatives of the energy which determines the best quadratic approximation of the energy to look for minimum energy states. Once converged to a minimum energy state, the surface may be analyzed for deformation profiles.\\[1ex]

\section{Binding transition (${W_\mathrm{1}}$)}
\label{A1}
The binding transition for particles to a membrane is determined by the mean curvature
of the particle at the contact point and is independent of surface tension \cite{deserno04}. To calculate the binding transition for any particle shape, the Monge parametrization can be used, where the surface is described by a height field, $h(\rhobf )$, where $\rhobf=(x,y) $ are the coordinates in the reference plane. For an almost planar membrane, a small-gradient approximation for the bending energy gives  
\begin{equation}
\mathcal{E_{\textrm{def }}} =  \int d A  \frac{\kappa}{2}  ( \nabla^2 h(\rhobf))^2    \, ,
\end{equation}
with $\int d A$ the integral over the reference plane. 

The critical adhesion strength at the binding transition marks the onset of adhesion, separating the unwrapped regime from the (partially) wrapped regime. Because it is completely determined by the competition between adhesion energy and bending energy, at the transition the mean curvature must equal the adhesion strength $\tilde w_{1}$,
\begin{equation}
\tilde w_1  =    a^2  \left[ \nabla^2 h(\rhobf=0) \right]^2   \, ,
\end{equation}
for the contact point between membrane and particle at $\rhobf=0$. For spheres $\left[ \nabla^2 h(\rhobf=0) \right]^2 = 4/a^2$, we find $\tilde w_{1}=4$, for prolate ellipsoids of aspect ratio b/a, such that the they are wrapped with their long axis parallel to the membrane, 
\begin{equation} \label{eq:W1PE}
\tilde w_1 (\textrm{PE}) = \left[ 1 + \left(\frac{a}{b}\right)^2 \right]^2 \, ,
\end{equation}
and for oblate ellipsoids,   
\begin{equation} \label{eq:W1OE}
\tilde w_{1}(\textrm{OE}) = 4 (b/a)^2 \, .
\end{equation}

\bibliographystyle{apsrev4-1}
\bibliography{references}

\begin{thebibliography}{71}%
\makeatletter
\providecommand \@ifxundefined [1]{%
 \@ifx{#1\undefined}
}%
\providecommand \@ifnum [1]{%
 \ifnum #1\expandafter \@firstoftwo
 \else \expandafter \@secondoftwo
 \fi
}%
\providecommand \@ifx [1]{%
 \ifx #1\expandafter \@firstoftwo
 \else \expandafter \@secondoftwo
 \fi
}%
\providecommand \natexlab [1]{#1}%
\providecommand \enquote  [1]{``#1''}%
\providecommand \bibnamefont  [1]{#1}%
\providecommand \bibfnamefont [1]{#1}%
\providecommand \citenamefont [1]{#1}%
\providecommand \href@noop [0]{\@secondoftwo}%
\providecommand \href [0]{\begingroup \@sanitize@url \@href}%
\providecommand \@href[1]{\@@startlink{#1}\@@href}%
\providecommand \@@href[1]{\endgroup#1\@@endlink}%
\providecommand \@sanitize@url [0]{\catcode `\\12\catcode `\$12\catcode
  `\&12\catcode `\#12\catcode `\^12\catcode `\_12\catcode `\%12\relax}%
\providecommand \@@startlink[1]{}%
\providecommand \@@endlink[0]{}%
\providecommand \url  [0]{\begingroup\@sanitize@url \@url }%
\providecommand \@url [1]{\endgroup\@href {#1}{\urlprefix }}%
\providecommand \urlprefix  [0]{URL }%
\providecommand \Eprint [0]{\href }%
\providecommand \doibase [0]{http://dx.doi.org/}%
\providecommand \selectlanguage [0]{\@gobble}%
\providecommand \bibinfo  [0]{\@secondoftwo}%
\providecommand \bibfield  [0]{\@secondoftwo}%
\providecommand \translation [1]{[#1]}%
\providecommand \BibitemOpen [0]{}%
\providecommand \bibitemStop [0]{}%
\providecommand \bibitemNoStop [0]{.\EOS\space}%
\providecommand \EOS [0]{\spacefactor3000\relax}%
\providecommand \BibitemShut  [1]{\csname bibitem#1\endcsname}%
\let\auto@bib@innerbib\@empty
\bibitem [{\citenamefont {McMahon}\ and\ \citenamefont
  {Gallop}(2005)}]{mcmahon05}%
  \BibitemOpen
  \bibfield  {author} {\bibinfo {author} {\bibfnamefont {H.~T.}\ \bibnamefont
  {McMahon}}\ and\ \bibinfo {author} {\bibfnamefont {J.~L.}\ \bibnamefont
  {Gallop}},\ }\href@noop {} {\bibfield  {journal} {\bibinfo  {journal}
  {Nature}\ }\textbf {\bibinfo {volume} {438}},\ \bibinfo {pages} {590}
  (\bibinfo {year} {2005})}\BibitemShut {NoStop}%
\bibitem [{\citenamefont {Hurley}\ \emph {et~al.}(2010)\citenamefont {Hurley},
  \citenamefont {Boura}, \citenamefont {Carlson},\ and\ \citenamefont
  {Różycki}}]{Hurley_2010}%
  \BibitemOpen
  \bibfield  {author} {\bibinfo {author} {\bibfnamefont {J.~H.}\ \bibnamefont
  {Hurley}}, \bibinfo {author} {\bibfnamefont {E.}~\bibnamefont {Boura}},
  \bibinfo {author} {\bibfnamefont {L.-A.}\ \bibnamefont {Carlson}}, \ and\
  \bibinfo {author} {\bibfnamefont {B.}~\bibnamefont {Różycki}},\ }\href@noop
  {} {\bibfield  {journal} {\bibinfo  {journal} {Cell}\ }\textbf {\bibinfo
  {volume} {143}},\ \bibinfo {pages} {875} (\bibinfo {year}
  {2010})}\BibitemShut {NoStop}%
\bibitem [{\citenamefont {Canton}\ and\ \citenamefont
  {Battaglia}(2012)}]{Canton_Battaglia_2012}%
  \BibitemOpen
  \bibfield  {author} {\bibinfo {author} {\bibfnamefont {I.}~\bibnamefont
  {Canton}}\ and\ \bibinfo {author} {\bibfnamefont {G.}~\bibnamefont
  {Battaglia}},\ }\href@noop {} {\bibfield  {journal} {\bibinfo  {journal}
  {Chem. Soc. Rev.}\ }\textbf {\bibinfo {volume} {41}},\ \bibinfo {pages}
  {2718} (\bibinfo {year} {2012})}\BibitemShut {NoStop}%
\bibitem [{\citenamefont {Kumar}\ \emph {et~al.}(1999)\citenamefont {Kumar},
  \citenamefont {Gompper},\ and\ \citenamefont {Lipowsky}}]{sunilkumar99}%
  \BibitemOpen
  \bibfield  {author} {\bibinfo {author} {\bibfnamefont {P.~B.~S.}\
  \bibnamefont {Kumar}}, \bibinfo {author} {\bibfnamefont {G.}~\bibnamefont
  {Gompper}}, \ and\ \bibinfo {author} {\bibfnamefont {R.}~\bibnamefont
  {Lipowsky}},\ }\href {\doibase 10.1103/PhysRevE.60.4610} {\bibfield
  {journal} {\bibinfo  {journal} {Phys. Rev. E}\ }\textbf {\bibinfo {volume}
  {60}},\ \bibinfo {pages} {4610} (\bibinfo {year} {1999})}\BibitemShut
  {NoStop}%
\bibitem [{\citenamefont {Sunil~Kumar}\ \emph {et~al.}(2001)\citenamefont
  {Sunil~Kumar}, \citenamefont {Gompper},\ and\ \citenamefont
  {Lipowsky}}]{sunilkumar01}%
  \BibitemOpen
  \bibfield  {author} {\bibinfo {author} {\bibfnamefont {P.~B.}\ \bibnamefont
  {Sunil~Kumar}}, \bibinfo {author} {\bibfnamefont {G.}~\bibnamefont
  {Gompper}}, \ and\ \bibinfo {author} {\bibfnamefont {R.}~\bibnamefont
  {Lipowsky}},\ }\href@noop {} {\bibfield  {journal} {\bibinfo  {journal}
  {Phys. Rev. Lett.}\ }\textbf {\bibinfo {volume} {86}},\ \bibinfo {pages}
  {3911} (\bibinfo {year} {2001})}\BibitemShut {NoStop}%
\bibitem [{\citenamefont {Baumgart}\ \emph {et~al.}(2003)\citenamefont
  {Baumgart}, \citenamefont {Hess},\ and\ \citenamefont {Webb}}]{baumgart03}%
  \BibitemOpen
  \bibfield  {author} {\bibinfo {author} {\bibfnamefont {T.}~\bibnamefont
  {Baumgart}}, \bibinfo {author} {\bibfnamefont {S.~T.}\ \bibnamefont {Hess}},
  \ and\ \bibinfo {author} {\bibfnamefont {W.~W.}\ \bibnamefont {Webb}},\
  }\href@noop {} {\bibfield  {journal} {\bibinfo  {journal} {Nature}\ }\textbf
  {\bibinfo {volume} {425}},\ \bibinfo {pages} {821} (\bibinfo {year}
  {2003})}\BibitemShut {NoStop}%
\bibitem [{\citenamefont {Semrau}\ \emph {et~al.}(2008)\citenamefont {Semrau},
  \citenamefont {Idema}, \citenamefont {Holtzer}, \citenamefont {Schmidt},\
  and\ \citenamefont {Storm}}]{semrau08}%
  \BibitemOpen
  \bibfield  {author} {\bibinfo {author} {\bibfnamefont {S.}~\bibnamefont
  {Semrau}}, \bibinfo {author} {\bibfnamefont {T.}~\bibnamefont {Idema}},
  \bibinfo {author} {\bibfnamefont {L.}~\bibnamefont {Holtzer}}, \bibinfo
  {author} {\bibfnamefont {T.}~\bibnamefont {Schmidt}}, \ and\ \bibinfo
  {author} {\bibfnamefont {C.}~\bibnamefont {Storm}},\ }\href@noop {}
  {\bibfield  {journal} {\bibinfo  {journal} {Phys. Rev. Lett.}\ }\textbf
  {\bibinfo {volume} {100}},\ \bibinfo {pages} {088101} (\bibinfo {year}
  {2008})}\BibitemShut {NoStop}%
\bibitem [{\citenamefont {Kohyama}\ \emph {et~al.}(2003)\citenamefont
  {Kohyama}, \citenamefont {Kroll},\ and\ \citenamefont
  {Gompper}}]{Kohyama_Kroll_Gompper03}%
  \BibitemOpen
  \bibfield  {author} {\bibinfo {author} {\bibfnamefont {T.}~\bibnamefont
  {Kohyama}}, \bibinfo {author} {\bibfnamefont {D.~M.}\ \bibnamefont {Kroll}},
  \ and\ \bibinfo {author} {\bibfnamefont {G.}~\bibnamefont {Gompper}},\
  }\href@noop {} {\bibfield  {journal} {\bibinfo  {journal} {Phys. Rev. E}\
  }\textbf {\bibinfo {volume} {68}},\ \bibinfo {pages} {061905} (\bibinfo
  {year} {2003})}\BibitemShut {NoStop}%
\bibitem [{\citenamefont {Reynwar}\ \emph {et~al.}(2007)\citenamefont
  {Reynwar}, \citenamefont {Illya}, \citenamefont {Harmandaris}, \citenamefont
  {Mueller}, \citenamefont {Kremer},\ and\ \citenamefont
  {Deserno}}]{reynwar07}%
  \BibitemOpen
  \bibfield  {author} {\bibinfo {author} {\bibfnamefont {B.~J.}\ \bibnamefont
  {Reynwar}}, \bibinfo {author} {\bibfnamefont {G.}~\bibnamefont {Illya}},
  \bibinfo {author} {\bibfnamefont {V.~A.}\ \bibnamefont {Harmandaris}},
  \bibinfo {author} {\bibfnamefont {M.~M.}\ \bibnamefont {Mueller}}, \bibinfo
  {author} {\bibfnamefont {K.}~\bibnamefont {Kremer}}, \ and\ \bibinfo {author}
  {\bibfnamefont {M.}~\bibnamefont {Deserno}},\ }\href@noop {} {\bibfield
  {journal} {\bibinfo  {journal} {Nature}\ }\textbf {\bibinfo {volume} {447}},\
  \bibinfo {pages} {461} (\bibinfo {year} {2007})}\BibitemShut {NoStop}%
\bibitem [{\citenamefont {Römer}\ \emph {et~al.}(2007)\citenamefont {Römer},
  \citenamefont {Berland}, \citenamefont {Chambon}, \citenamefont {Gaus},
  \citenamefont {Windschiegl}, \citenamefont {Tenza}, \citenamefont {Aly},
  \citenamefont {Fraisier}, \citenamefont {Florent}, \citenamefont {Perrais},\
  and\ \citenamefont {et~al.}}]{Bassereau_2007}%
  \BibitemOpen
  \bibfield  {author} {\bibinfo {author} {\bibfnamefont {W.}~\bibnamefont
  {Römer}}, \bibinfo {author} {\bibfnamefont {L.}~\bibnamefont {Berland}},
  \bibinfo {author} {\bibfnamefont {V.}~\bibnamefont {Chambon}}, \bibinfo
  {author} {\bibfnamefont {K.}~\bibnamefont {Gaus}}, \bibinfo {author}
  {\bibfnamefont {B.}~\bibnamefont {Windschiegl}}, \bibinfo {author}
  {\bibfnamefont {D.}~\bibnamefont {Tenza}}, \bibinfo {author} {\bibfnamefont
  {M.~R.~E.}\ \bibnamefont {Aly}}, \bibinfo {author} {\bibfnamefont
  {V.}~\bibnamefont {Fraisier}}, \bibinfo {author} {\bibfnamefont {J.~C.}\
  \bibnamefont {Florent}}, \bibinfo {author} {\bibfnamefont {D.}~\bibnamefont
  {Perrais}}, \ and\ \bibinfo {author} {\bibnamefont {et~al.}},\ }\href@noop {}
  {\bibfield  {journal} {\bibinfo  {journal} {Nature}\ }\textbf {\bibinfo
  {volume} {450}},\ \bibinfo {pages} {670} (\bibinfo {year}
  {2007})}\BibitemShut {NoStop}%
\bibitem [{\citenamefont {Sens}\ \emph {et~al.}(2008)\citenamefont {Sens},
  \citenamefont {Johannes},\ and\ \citenamefont {Bassereau}}]{sens08}%
  \BibitemOpen
  \bibfield  {author} {\bibinfo {author} {\bibfnamefont {P.}~\bibnamefont
  {Sens}}, \bibinfo {author} {\bibfnamefont {L.}~\bibnamefont {Johannes}}, \
  and\ \bibinfo {author} {\bibfnamefont {P.}~\bibnamefont {Bassereau}},\
  }\href@noop {} {\bibfield  {journal} {\bibinfo  {journal} {Curr. Opin. Cell
  Biol.}\ }\textbf {\bibinfo {volume} {20}},\ \bibinfo {pages} {476} (\bibinfo
  {year} {2008})}\BibitemShut {NoStop}%
\bibitem [{\citenamefont {Auth}\ and\ \citenamefont {Gompper}(2009)}]{auth09}%
  \BibitemOpen
  \bibfield  {author} {\bibinfo {author} {\bibfnamefont {T.}~\bibnamefont
  {Auth}}\ and\ \bibinfo {author} {\bibfnamefont {G.}~\bibnamefont {Gompper}},\
  }\href@noop {} {\bibfield  {journal} {\bibinfo  {journal} {Phys. Rev. E.}\
  }\textbf {\bibinfo {volume} {80}},\ \bibinfo {pages} {031901} (\bibinfo
  {year} {2009})}\BibitemShut {NoStop}%
\bibitem [{\citenamefont {Lipowsky}(1992)}]{Lipowsky92}%
  \BibitemOpen
  \bibfield  {author} {\bibinfo {author} {\bibfnamefont {R.}~\bibnamefont
  {Lipowsky}},\ }\href@noop {} {\bibfield  {journal} {\bibinfo  {journal} {J.
  Phys. II (France)}\ }\textbf {\bibinfo {volume} {2}},\ \bibinfo {pages}
  {1825} (\bibinfo {year} {1992})}\BibitemShut {NoStop}%
\bibitem [{\citenamefont {Lipowsky}(1993)}]{Lipowsky93}%
  \BibitemOpen
  \bibfield  {author} {\bibinfo {author} {\bibfnamefont {R.}~\bibnamefont
  {Lipowsky}},\ }\href@noop {} {\bibfield  {journal} {\bibinfo  {journal}
  {Biophys. J.}\ }\textbf {\bibinfo {volume} {64}},\ \bibinfo {pages} {1133}
  (\bibinfo {year} {1993})}\BibitemShut {NoStop}%
\bibitem [{\citenamefont {Jülicher}\ and\ \citenamefont
  {Lipowsky}(1993)}]{Juelicher93}%
  \BibitemOpen
  \bibfield  {author} {\bibinfo {author} {\bibfnamefont {F.}~\bibnamefont
  {Jülicher}}\ and\ \bibinfo {author} {\bibfnamefont {R.}~\bibnamefont
  {Lipowsky}},\ }\href@noop {} {\bibfield  {journal} {\bibinfo  {journal}
  {Phys. Rev. Lett.}\ }\textbf {\bibinfo {volume} {70}},\ \bibinfo {pages}
  {2964} (\bibinfo {year} {1993})}\BibitemShut {NoStop}%
\bibitem [{\citenamefont {Champion}\ and\ \citenamefont
  {Mitragotri}(2006)}]{Champion_Mitragotri06}%
  \BibitemOpen
  \bibfield  {author} {\bibinfo {author} {\bibfnamefont {J.~A.}\ \bibnamefont
  {Champion}}\ and\ \bibinfo {author} {\bibfnamefont {S.}~\bibnamefont
  {Mitragotri}},\ }\href@noop {} {\bibfield  {journal} {\bibinfo  {journal}
  {Proc. Natl. Acad. Sci. U.S.A.}\ }\textbf {\bibinfo {volume} {103}},\
  \bibinfo {pages} {4930} (\bibinfo {year} {2006})}\BibitemShut {NoStop}%
\bibitem [{\citenamefont {Chithrani}\ \emph {et~al.}(2006)\citenamefont
  {Chithrani}, \citenamefont {Ghazani},\ and\ \citenamefont
  {Chan}}]{Chithrani06}%
  \BibitemOpen
  \bibfield  {author} {\bibinfo {author} {\bibfnamefont {B.~D.}\ \bibnamefont
  {Chithrani}}, \bibinfo {author} {\bibfnamefont {A.~A.}\ \bibnamefont
  {Ghazani}}, \ and\ \bibinfo {author} {\bibfnamefont {W.~C.~W.}\ \bibnamefont
  {Chan}},\ }\href@noop {} {\bibfield  {journal} {\bibinfo  {journal} {Nano
  Lett.}\ }\textbf {\bibinfo {volume} {6}},\ \bibinfo {pages} {662} (\bibinfo
  {year} {2006})}\BibitemShut {NoStop}%
\bibitem [{\citenamefont {Decuzzi}\ \emph {et~al.}(2009)\citenamefont
  {Decuzzi}, \citenamefont {Pasqualini}, \citenamefont {Arap},\ and\
  \citenamefont {Ferrari}}]{Decuzzi_Ferrari_2009}%
  \BibitemOpen
  \bibfield  {author} {\bibinfo {author} {\bibfnamefont {P.}~\bibnamefont
  {Decuzzi}}, \bibinfo {author} {\bibfnamefont {R.}~\bibnamefont {Pasqualini}},
  \bibinfo {author} {\bibfnamefont {W.}~\bibnamefont {Arap}}, \ and\ \bibinfo
  {author} {\bibfnamefont {M.}~\bibnamefont {Ferrari}},\ }\href@noop {}
  {\bibfield  {journal} {\bibinfo  {journal} {Pharm. Res.}\ }\textbf {\bibinfo
  {volume} {26}},\ \bibinfo {pages} {235} (\bibinfo {year} {2009})}\BibitemShut
  {NoStop}%
\bibitem [{\citenamefont {Best}\ \emph {et~al.}(2012)\citenamefont {Best},
  \citenamefont {Yan},\ and\ \citenamefont {Caruso}}]{Frank12}%
  \BibitemOpen
  \bibfield  {author} {\bibinfo {author} {\bibfnamefont {J.~P.}\ \bibnamefont
  {Best}}, \bibinfo {author} {\bibfnamefont {Y.}~\bibnamefont {Yan}}, \ and\
  \bibinfo {author} {\bibfnamefont {F.}~\bibnamefont {Caruso}},\ }\href
  {\doibase 10.1002/adhm.201100012} {\bibfield  {journal} {\bibinfo  {journal}
  {Adv. Healthcare Mater.}\ }\textbf {\bibinfo {volume} {1}},\ \bibinfo {pages}
  {35} (\bibinfo {year} {2012})}\BibitemShut {NoStop}%
\bibitem [{\citenamefont {Florez}\ \emph {et~al.}(2012)\citenamefont {Florez},
  \citenamefont {Herrmann}, \citenamefont {Cramer}, \citenamefont {Hauser},
  \citenamefont {Koynov}, \citenamefont {Landfester}, \citenamefont {Crespy},\
  and\ \citenamefont {Mailänder}}]{Florez_landfester12}%
  \BibitemOpen
  \bibfield  {author} {\bibinfo {author} {\bibfnamefont {L.}~\bibnamefont
  {Florez}}, \bibinfo {author} {\bibfnamefont {C.}~\bibnamefont {Herrmann}},
  \bibinfo {author} {\bibfnamefont {J.~M.}\ \bibnamefont {Cramer}}, \bibinfo
  {author} {\bibfnamefont {C.~P.}\ \bibnamefont {Hauser}}, \bibinfo {author}
  {\bibfnamefont {K.}~\bibnamefont {Koynov}}, \bibinfo {author} {\bibfnamefont
  {K.}~\bibnamefont {Landfester}}, \bibinfo {author} {\bibfnamefont
  {D.}~\bibnamefont {Crespy}}, \ and\ \bibinfo {author} {\bibfnamefont
  {V.}~\bibnamefont {Mailänder}},\ }\href {\doibase 10.1002/smll.201102002}
  {\bibfield  {journal} {\bibinfo  {journal} {Small}\ }\textbf {\bibinfo
  {volume} {8}},\ \bibinfo {pages} {2222} (\bibinfo {year} {2012})}\BibitemShut
  {NoStop}%
\bibitem [{\citenamefont {Tzlil}\ \emph {et~al.}(2004)\citenamefont {Tzlil},
  \citenamefont {Deserno}, \citenamefont {Gelbart},\ and\ \citenamefont
  {Ben-Shaul}}]{tzlil04}%
  \BibitemOpen
  \bibfield  {author} {\bibinfo {author} {\bibfnamefont {S.}~\bibnamefont
  {Tzlil}}, \bibinfo {author} {\bibfnamefont {M.}~\bibnamefont {Deserno}},
  \bibinfo {author} {\bibfnamefont {W.~M.}\ \bibnamefont {Gelbart}}, \ and\
  \bibinfo {author} {\bibfnamefont {A.}~\bibnamefont {Ben-Shaul}},\ }\href@noop
  {} {\bibfield  {journal} {\bibinfo  {journal} {Biophys. J.}\ }\textbf
  {\bibinfo {volume} {86}},\ \bibinfo {pages} {2037} (\bibinfo {year}
  {2004})}\BibitemShut {NoStop}%
\bibitem [{\citenamefont {Mercer}\ \emph {et~al.}(2010)\citenamefont {Mercer},
  \citenamefont {Schelhaas},\ and\ \citenamefont
  {Helenius}}]{Mercer_Helenius_2010}%
  \BibitemOpen
  \bibfield  {author} {\bibinfo {author} {\bibfnamefont {J.}~\bibnamefont
  {Mercer}}, \bibinfo {author} {\bibfnamefont {M.}~\bibnamefont {Schelhaas}}, \
  and\ \bibinfo {author} {\bibfnamefont {A.}~\bibnamefont {Helenius}},\
  }\href@noop {} {\bibfield  {journal} {\bibinfo  {journal} {Annu. Rev.
  Biochem.}\ }\textbf {\bibinfo {volume} {79}},\ \bibinfo {pages} {803}
  (\bibinfo {year} {2010})}\BibitemShut {NoStop}%
\bibitem [{\citenamefont {Gratton}\ \emph {et~al.}(2008)\citenamefont
  {Gratton}, \citenamefont {Ropp}, \citenamefont {Pohlhaus}, \citenamefont
  {Luft}, \citenamefont {Madden}, \citenamefont {Napier},\ and\ \citenamefont
  {DeSimone}}]{Gratton_DeSimone_2008}%
  \BibitemOpen
  \bibfield  {author} {\bibinfo {author} {\bibfnamefont {S.~E.~A.}\
  \bibnamefont {Gratton}}, \bibinfo {author} {\bibfnamefont {P.~A.}\
  \bibnamefont {Ropp}}, \bibinfo {author} {\bibfnamefont {P.~D.}\ \bibnamefont
  {Pohlhaus}}, \bibinfo {author} {\bibfnamefont {J.~C.}\ \bibnamefont {Luft}},
  \bibinfo {author} {\bibfnamefont {V.~J.}\ \bibnamefont {Madden}}, \bibinfo
  {author} {\bibfnamefont {M.~E.}\ \bibnamefont {Napier}}, \ and\ \bibinfo
  {author} {\bibfnamefont {J.~M.}\ \bibnamefont {DeSimone}},\ }\href@noop {}
  {\bibfield  {journal} {\bibinfo  {journal} {Proc. Natl. Acad. Sci. U.S.A.}\
  }\textbf {\bibinfo {volume} {105}},\ \bibinfo {pages} {11613} (\bibinfo
  {year} {2008})}\BibitemShut {NoStop}%
\bibitem [{\citenamefont {Barua}\ \emph {et~al.}(2013)\citenamefont {Barua},
  \citenamefont {Yoo}, \citenamefont {Kolhar}, \citenamefont {Wakankar},
  \citenamefont {Gokarn},\ and\ \citenamefont {Mitragotri}}]{barua13}%
  \BibitemOpen
  \bibfield  {author} {\bibinfo {author} {\bibfnamefont {S.}~\bibnamefont
  {Barua}}, \bibinfo {author} {\bibfnamefont {J.-W.}\ \bibnamefont {Yoo}},
  \bibinfo {author} {\bibfnamefont {P.}~\bibnamefont {Kolhar}}, \bibinfo
  {author} {\bibfnamefont {A.}~\bibnamefont {Wakankar}}, \bibinfo {author}
  {\bibfnamefont {Y.~R.}\ \bibnamefont {Gokarn}}, \ and\ \bibinfo {author}
  {\bibfnamefont {S.}~\bibnamefont {Mitragotri}},\ }\href@noop {} {\bibfield
  {journal} {\bibinfo  {journal} {Proc. Natl. Acad. Sci. U.S.A.}\ }\textbf
  {\bibinfo {volume} {110}},\ \bibinfo {pages} {3270} (\bibinfo {year}
  {2013})}\BibitemShut {NoStop}%
\bibitem [{\citenamefont {Kubo}\ \emph {et~al.}(2009)\citenamefont {Kubo},
  \citenamefont {Freitas-Astúa}, \citenamefont {Machado},\ and\ \citenamefont
  {Kitajima}}]{Kubo_2009}%
  \BibitemOpen
  \bibfield  {author} {\bibinfo {author} {\bibfnamefont {K.}~\bibnamefont
  {Kubo}}, \bibinfo {author} {\bibfnamefont {J.}~\bibnamefont
  {Freitas-Astúa}}, \bibinfo {author} {\bibfnamefont {M.}~\bibnamefont
  {Machado}}, \ and\ \bibinfo {author} {\bibfnamefont {E.}~\bibnamefont
  {Kitajima}},\ }\href {\doibase 10.1007/s10327-009-0167-z} {\bibfield
  {journal} {\bibinfo  {journal} {J. Gen. Plant Pathol.}\ }\textbf {\bibinfo
  {volume} {75}},\ \bibinfo {pages} {250} (\bibinfo {year} {2009})}\BibitemShut
  {NoStop}%
\bibitem [{\citenamefont {Liu}\ \emph {et~al.}(2012)\citenamefont {Liu},
  \citenamefont {Tan}, \citenamefont {Thomas}, \citenamefont {Ou-Yang},\ and\
  \citenamefont {Muzykantov}}]{liu2012}%
  \BibitemOpen
  \bibfield  {author} {\bibinfo {author} {\bibfnamefont {Y.}~\bibnamefont
  {Liu}}, \bibinfo {author} {\bibfnamefont {J.}~\bibnamefont {Tan}}, \bibinfo
  {author} {\bibfnamefont {A.}~\bibnamefont {Thomas}}, \bibinfo {author}
  {\bibfnamefont {D.}~\bibnamefont {Ou-Yang}}, \ and\ \bibinfo {author}
  {\bibfnamefont {V.}~\bibnamefont {Muzykantov}},\ }\href@noop {} {\bibfield
  {journal} {\bibinfo  {journal} {Ther. Delivery}\ }\textbf {\bibinfo {volume}
  {3}},\ \bibinfo {pages} {181} (\bibinfo {year} {2012})}\BibitemShut {NoStop}%
\bibitem [{\citenamefont {Pissuwan}\ \emph {et~al.}(2008)\citenamefont
  {Pissuwan}, \citenamefont {Valenzuela},\ and\ \citenamefont
  {Cortie}}]{Pissuwan_Cortie_2008}%
  \BibitemOpen
  \bibfield  {author} {\bibinfo {author} {\bibfnamefont {D.}~\bibnamefont
  {Pissuwan}}, \bibinfo {author} {\bibfnamefont {S.}~\bibnamefont
  {Valenzuela}}, \ and\ \bibinfo {author} {\bibfnamefont {M.~B.}\ \bibnamefont
  {Cortie}},\ }\href@noop {} {\bibfield  {journal} {\bibinfo  {journal}
  {Biotechnol. Genet. Eng. Rev.}\ }\textbf {\bibinfo {volume} {25}},\ \bibinfo
  {pages} {93} (\bibinfo {year} {2008})}\BibitemShut {NoStop}%
\bibitem [{\citenamefont {Xu}\ \emph {et~al.}(2011)\citenamefont {Xu},
  \citenamefont {Kuang}, \citenamefont {Wang},\ and\ \citenamefont
  {Xu}}]{Xu_2011}%
  \BibitemOpen
  \bibfield  {author} {\bibinfo {author} {\bibfnamefont {L.}~\bibnamefont
  {Xu}}, \bibinfo {author} {\bibfnamefont {H.}~\bibnamefont {Kuang}}, \bibinfo
  {author} {\bibfnamefont {L.}~\bibnamefont {Wang}}, \ and\ \bibinfo {author}
  {\bibfnamefont {C.}~\bibnamefont {Xu}},\ }\href@noop {} {\bibfield  {journal}
  {\bibinfo  {journal} {J. Mater. Chem.}\ }\textbf {\bibinfo {volume} {21}},\
  \bibinfo {pages} {16759} (\bibinfo {year} {2011})}\BibitemShut {NoStop}%
\bibitem [{\citenamefont {Dietrich}\ \emph {et~al.}(1997)\citenamefont
  {Dietrich}, \citenamefont {Angelova},\ and\ \citenamefont
  {Pouligny}}]{Dietrich_Angelova_Pouligny_1997}%
  \BibitemOpen
  \bibfield  {author} {\bibinfo {author} {\bibfnamefont {C.}~\bibnamefont
  {Dietrich}}, \bibinfo {author} {\bibfnamefont {M.}~\bibnamefont {Angelova}},
  \ and\ \bibinfo {author} {\bibfnamefont {B.}~\bibnamefont {Pouligny}},\
  }\href@noop {} {\bibfield  {journal} {\bibinfo  {journal} {J. Phys. II
  (France)}\ }\textbf {\bibinfo {volume} {7}},\ \bibinfo {pages} {1651}
  (\bibinfo {year} {1997})}\BibitemShut {NoStop}%
\bibitem [{\citenamefont {Lipowsky}\ and\ \citenamefont
  {Döbereiner}(1998)}]{Lipowsky_Dobereiner_1998}%
  \BibitemOpen
  \bibfield  {author} {\bibinfo {author} {\bibfnamefont {R.}~\bibnamefont
  {Lipowsky}}\ and\ \bibinfo {author} {\bibfnamefont {H.~G.}\ \bibnamefont
  {Döbereiner}},\ }\href@noop {} {\bibfield  {journal} {\bibinfo  {journal}
  {Europhys. Lett.}\ }\textbf {\bibinfo {volume} {43}},\ \bibinfo {pages} {219}
  (\bibinfo {year} {1998})}\BibitemShut {NoStop}%
\bibitem [{\citenamefont {Zhang}\ \emph
  {et~al.}(2012{\natexlab{a}})\citenamefont {Zhang}, \citenamefont {Nelson},\
  and\ \citenamefont {Beales}}]{Zhang_Beales_2012}%
  \BibitemOpen
  \bibfield  {author} {\bibinfo {author} {\bibfnamefont {S.}~\bibnamefont
  {Zhang}}, \bibinfo {author} {\bibfnamefont {A.}~\bibnamefont {Nelson}}, \
  and\ \bibinfo {author} {\bibfnamefont {P.~A.}\ \bibnamefont {Beales}},\
  }\href@noop {} {\bibfield  {journal} {\bibinfo  {journal} {Langmuir}\
  }\textbf {\bibinfo {volume} {28}},\ \bibinfo {pages} {12831} (\bibinfo {year}
  {2012}{\natexlab{a}})}\BibitemShut {NoStop}%
\bibitem [{\citenamefont {Benoit}\ and\ \citenamefont
  {Saxena}(2007)}]{Benoit_Saxena_2007}%
  \BibitemOpen
  \bibfield  {author} {\bibinfo {author} {\bibfnamefont {J.}~\bibnamefont
  {Benoit}}\ and\ \bibinfo {author} {\bibfnamefont {A.}~\bibnamefont
  {Saxena}},\ }\href@noop {} {\bibfield  {journal} {\bibinfo  {journal} {Phys.
  Rev. E}\ }\textbf {\bibinfo {volume} {76}},\ \bibinfo {pages} {041912}
  (\bibinfo {year} {2007})}\BibitemShut {NoStop}%
\bibitem [{\citenamefont {Deserno}\ and\ \citenamefont
  {Gelbart}(2002)}]{deserno02}%
  \BibitemOpen
  \bibfield  {author} {\bibinfo {author} {\bibfnamefont {M.}~\bibnamefont
  {Deserno}}\ and\ \bibinfo {author} {\bibfnamefont {W.}~\bibnamefont
  {Gelbart}},\ }\href@noop {} {\bibfield  {journal} {\bibinfo  {journal} {J.
  Phys. Chem. B}\ }\textbf {\bibinfo {volume} {106}},\ \bibinfo {pages} {5543}
  (\bibinfo {year} {2002})}\BibitemShut {NoStop}%
\bibitem [{\citenamefont {Deserno}\ and\ \citenamefont
  {Bickel}(2003)}]{deserno03}%
  \BibitemOpen
  \bibfield  {author} {\bibinfo {author} {\bibfnamefont {M.}~\bibnamefont
  {Deserno}}\ and\ \bibinfo {author} {\bibfnamefont {T.}~\bibnamefont
  {Bickel}},\ }\href@noop {} {\bibfield  {journal} {\bibinfo  {journal}
  {Europhys. Lett.}\ }\textbf {\bibinfo {volume} {62}},\ \bibinfo {pages} {767}
  (\bibinfo {year} {2003})}\BibitemShut {NoStop}%
\bibitem [{\citenamefont {Deserno}(2003)}]{deserno04}%
  \BibitemOpen
  \bibfield  {author} {\bibinfo {author} {\bibfnamefont {M.}~\bibnamefont
  {Deserno}},\ }\href@noop {} {\bibfield  {journal} {\bibinfo  {journal} {Phys.
  Rev. E.}\ }\textbf {\bibinfo {volume} {69}},\ \bibinfo {pages} {031903}
  (\bibinfo {year} {2003})}\BibitemShut {NoStop}%
\bibitem [{\citenamefont {Deserno}(2004)}]{deserno04a}%
  \BibitemOpen
  \bibfield  {author} {\bibinfo {author} {\bibfnamefont {M.}~\bibnamefont
  {Deserno}},\ }\href@noop {} {\bibfield  {journal} {\bibinfo  {journal} {J.
  Phys.: Condens. Matter}\ }\textbf {\bibinfo {volume} {16}},\ \bibinfo {pages}
  {S2061–} (\bibinfo {year} {2004})}\BibitemShut {NoStop}%
\bibitem [{\citenamefont {Weikl}(2003)}]{Weikl03}%
  \BibitemOpen
  \bibfield  {author} {\bibinfo {author} {\bibfnamefont {T.~R.}\ \bibnamefont
  {Weikl}},\ }\href@noop {} {\bibfield  {journal} {\bibinfo  {journal} {Eur.
  Phys. J. E}\ }\textbf {\bibinfo {volume} {12}},\ \bibinfo {pages} {9}
  (\bibinfo {year} {2003})}\BibitemShut {NoStop}%
\bibitem [{\citenamefont {Mkrtchyan}\ \emph {et~al.}(2010)\citenamefont
  {Mkrtchyan}, \citenamefont {Ing},\ and\ \citenamefont
  {Chen}}]{Mkrtchyan_Chen10}%
  \BibitemOpen
  \bibfield  {author} {\bibinfo {author} {\bibfnamefont {S.}~\bibnamefont
  {Mkrtchyan}}, \bibinfo {author} {\bibfnamefont {C.}~\bibnamefont {Ing}}, \
  and\ \bibinfo {author} {\bibfnamefont {J.~Z.~Y.}\ \bibnamefont {Chen}},\
  }\href@noop {} {\bibfield  {journal} {\bibinfo  {journal} {Phys. Rev. E}\
  }\textbf {\bibinfo {volume} {81}},\ \bibinfo {pages} {1} (\bibinfo {year}
  {2010})}\BibitemShut {NoStop}%
\bibitem [{\citenamefont {Canham}(1970)}]{Canham1970}%
  \BibitemOpen
  \bibfield  {author} {\bibinfo {author} {\bibfnamefont {P.}~\bibnamefont
  {Canham}},\ }\href {\doibase 10.1016/S0022-5193(70)80032-7} {\bibfield
  {journal} {\bibinfo  {journal} {J. Theor. Biol.}\ }\textbf {\bibinfo {volume}
  {26}},\ \bibinfo {pages} {61 } (\bibinfo {year} {1970})}\BibitemShut
  {NoStop}%
\bibitem [{\citenamefont {Helfrich}(1973)}]{Helfrich_1973}%
  \BibitemOpen
  \bibfield  {author} {\bibinfo {author} {\bibfnamefont {W.}~\bibnamefont
  {Helfrich}},\ }\href@noop {} {\bibfield  {journal} {\bibinfo  {journal} {Z.
  Naturforsch. C}\ }\textbf {\bibinfo {volume} {28}},\ \bibinfo {pages} {693}
  (\bibinfo {year} {1973})}\BibitemShut {NoStop}%
\bibitem [{\citenamefont {Bloor}\ and\ \citenamefont
  {Wilson}(2000)}]{Bloor_Wilson00}%
  \BibitemOpen
  \bibfield  {author} {\bibinfo {author} {\bibfnamefont {M.~I.}\ \bibnamefont
  {Bloor}}\ and\ \bibinfo {author} {\bibfnamefont {M.~J.}\ \bibnamefont
  {Wilson}},\ }\href@noop {} {\bibfield  {journal} {\bibinfo  {journal} {Phys.
  Rev. E}\ }\textbf {\bibinfo {volume} {61}},\ \bibinfo {pages} {4218}
  (\bibinfo {year} {2000})}\BibitemShut {NoStop}%
\bibitem [{\citenamefont {Zhong-can}\ and\ \citenamefont
  {Helfrich}(1989)}]{OuYang_Helfrich_1989}%
  \BibitemOpen
  \bibfield  {author} {\bibinfo {author} {\bibfnamefont {O.-Y.}\ \bibnamefont
  {Zhong-can}}\ and\ \bibinfo {author} {\bibfnamefont {W.}~\bibnamefont
  {Helfrich}},\ }\href {\doibase 10.1103/PhysRevA.39.5280} {\bibfield
  {journal} {\bibinfo  {journal} {Phys. Rev. A}\ }\textbf {\bibinfo {volume}
  {39}},\ \bibinfo {pages} {5280} (\bibinfo {year} {1989})}\BibitemShut
  {NoStop}%
\bibitem [{\citenamefont {Seifert}\ \emph {et~al.}(1991)\citenamefont
  {Seifert}, \citenamefont {Berndl},\ and\ \citenamefont
  {Lipowsky}}]{Seifert_Berndl_1991}%
  \BibitemOpen
  \bibfield  {author} {\bibinfo {author} {\bibfnamefont {U.}~\bibnamefont
  {Seifert}}, \bibinfo {author} {\bibfnamefont {K.}~\bibnamefont {Berndl}}, \
  and\ \bibinfo {author} {\bibfnamefont {R.}~\bibnamefont {Lipowsky}},\ }\href
  {\doibase 10.1103/PhysRevA.44.1182} {\bibfield  {journal} {\bibinfo
  {journal} {Phys. Rev. A}\ }\textbf {\bibinfo {volume} {44}},\ \bibinfo
  {pages} {1182} (\bibinfo {year} {1991})}\BibitemShut {NoStop}%
\bibitem [{\citenamefont {Jülicher}\ and\ \citenamefont
  {Lipowsky}(1996)}]{Julicher_Lipowsky96}%
  \BibitemOpen
  \bibfield  {author} {\bibinfo {author} {\bibfnamefont {F.}~\bibnamefont
  {Jülicher}}\ and\ \bibinfo {author} {\bibfnamefont {R.}~\bibnamefont
  {Lipowsky}},\ }\href@noop {} {\bibfield  {journal} {\bibinfo  {journal}
  {Phys. Rev. E}\ }\textbf {\bibinfo {volume} {53}},\ \bibinfo {pages} {2670}
  (\bibinfo {year} {1996})}\BibitemShut {NoStop}%
\bibitem [{\citenamefont {Nowak}\ and\ \citenamefont {Chou}(2008)}]{nowak08}%
  \BibitemOpen
  \bibfield  {author} {\bibinfo {author} {\bibfnamefont {S.~A.}\ \bibnamefont
  {Nowak}}\ and\ \bibinfo {author} {\bibfnamefont {T.}~\bibnamefont {Chou}},\
  }\href@noop {} {\bibfield  {journal} {\bibinfo  {journal} {Phys. Rev. E.}\
  }\textbf {\bibinfo {volume} {78}},\ \bibinfo {pages} {021908} (\bibinfo
  {year} {2008})}\BibitemShut {NoStop}%
\bibitem [{\citenamefont {Khairy}\ and\ \citenamefont
  {Howard}(2011)}]{Khairy_Howard11}%
  \BibitemOpen
  \bibfield  {author} {\bibinfo {author} {\bibfnamefont {K.}~\bibnamefont
  {Khairy}}\ and\ \bibinfo {author} {\bibfnamefont {J.}~\bibnamefont
  {Howard}},\ }\href@noop {} {\bibfield  {journal} {\bibinfo  {journal} {Soft
  Matter}\ }\textbf {\bibinfo {volume} {7}},\ \bibinfo {pages} {2138} (\bibinfo
  {year} {2011})}\BibitemShut {NoStop}%
\bibitem [{\citenamefont {G{\'o}{\'z}d{\'z}}\ and\ \citenamefont
  {Gompper}(2001)}]{gozdz01}%
  \BibitemOpen
  \bibfield  {author} {\bibinfo {author} {\bibfnamefont {W.}~\bibnamefont
  {G{\'o}{\'z}d{\'z}}}\ and\ \bibinfo {author} {\bibfnamefont {G.}~\bibnamefont
  {Gompper}},\ }\href@noop {} {\bibfield  {journal} {\bibinfo  {journal}
  {Europhys. Lett.)}\ }\textbf {\bibinfo {volume} {55}},\ \bibinfo {pages}
  {587} (\bibinfo {year} {2001})}\BibitemShut {NoStop}%
\bibitem [{\citenamefont {Góźdź}(2007)}]{Gozdz_2007}%
  \BibitemOpen
  \bibfield  {author} {\bibinfo {author} {\bibfnamefont {W.~T.}\ \bibnamefont
  {Góźdź}},\ }\href@noop {} {\bibfield  {journal} {\bibinfo  {journal}
  {Langmuir}\ }\textbf {\bibinfo {volume} {23}},\ \bibinfo {pages} {5665}
  (\bibinfo {year} {2007})}\BibitemShut {NoStop}%
\bibitem [{\citenamefont {Brakke}(1992)}]{Brakke92}%
  \BibitemOpen
  \bibfield  {author} {\bibinfo {author} {\bibfnamefont {K.~A.}\ \bibnamefont
  {Brakke}},\ }\href {\doibase 10.1080/10586458.1992.10504253} {\bibfield
  {journal} {\bibinfo  {journal} {Exp. Math.}\ }\textbf {\bibinfo {volume}
  {1}},\ \bibinfo {pages} {141} (\bibinfo {year} {1992})}\BibitemShut {NoStop}%
\bibitem [{\citenamefont {Wintz}\ \emph {et~al.}(1996)\citenamefont {Wintz},
  \citenamefont {Döbereiner},\ and\ \citenamefont
  {Seifert}}]{Wintz_Dobereiner_Seifert_1996}%
  \BibitemOpen
  \bibfield  {author} {\bibinfo {author} {\bibfnamefont {W.}~\bibnamefont
  {Wintz}}, \bibinfo {author} {\bibfnamefont {H.~G.}\ \bibnamefont
  {Döbereiner}}, \ and\ \bibinfo {author} {\bibfnamefont {U.}~\bibnamefont
  {Seifert}},\ }\href@noop {} {\bibfield  {journal} {\bibinfo  {journal}
  {Europhys. Lett.}\ }\textbf {\bibinfo {volume} {33}},\ \bibinfo {pages} {403}
  (\bibinfo {year} {1996})}\BibitemShut {NoStop}%
\bibitem [{kan()}]{kantor1989}%
  \BibitemOpen
  \href@noop {} {\emph {\bibinfo {title} {G. Gompper and D. M. Kroll: in
  Statistical Mechanics of Membranes and Surfaces, ed. D. R. Nelson, T. Piran,
  and S. Weinberg (World Scientific, Singapore, 2004) 2nd ed.}}}\BibitemShut
  {Stop}%
\bibitem [{\citenamefont {Kroll}\ and\ \citenamefont
  {Gompper}(1992)}]{gompper92}%
  \BibitemOpen
  \bibfield  {author} {\bibinfo {author} {\bibfnamefont {D.~M.}\ \bibnamefont
  {Kroll}}\ and\ \bibinfo {author} {\bibfnamefont {G.}~\bibnamefont
  {Gompper}},\ }\href {\doibase 10.1126/science.1546294} {\bibfield  {journal}
  {\bibinfo  {journal} {Science}\ }\textbf {\bibinfo {volume} {255}},\ \bibinfo
  {pages} {968} (\bibinfo {year} {1992})}\BibitemShut {NoStop}%
\bibitem [{\citenamefont {Gompper}\ and\ \citenamefont
  {Kroll}(1997)}]{gompper1997}%
  \BibitemOpen
  \bibfield  {author} {\bibinfo {author} {\bibfnamefont {G.}~\bibnamefont
  {Gompper}}\ and\ \bibinfo {author} {\bibfnamefont {D.~M.}\ \bibnamefont
  {Kroll}},\ }\href@noop {} {\bibfield  {journal} {\bibinfo  {journal} {J.
  Phys.: Condens. Matter}\ }\textbf {\bibinfo {volume} {9}},\ \bibinfo {pages}
  {8795} (\bibinfo {year} {1997})}\BibitemShut {NoStop}%
\bibitem [{\citenamefont {Fošnarič}\ \emph {et~al.}(2009)\citenamefont
  {Fošnarič}, \citenamefont {Iglič}, \citenamefont {Kroll},\ and\
  \citenamefont {May}}]{fosnaric09}%
  \BibitemOpen
  \bibfield  {author} {\bibinfo {author} {\bibfnamefont {M.}~\bibnamefont
  {Fošnarič}}, \bibinfo {author} {\bibfnamefont {A.}~\bibnamefont {Iglič}},
  \bibinfo {author} {\bibfnamefont {D.~M.}\ \bibnamefont {Kroll}}, \ and\
  \bibinfo {author} {\bibfnamefont {S.}~\bibnamefont {May}},\ }\href@noop {}
  {\bibfield  {journal} {\bibinfo  {journal} {J. Chem. Phys.}\ }\textbf
  {\bibinfo {volume} {131}},\ \bibinfo {pages} {105103} (\bibinfo {year}
  {2009})}\BibitemShut {NoStop}%
\bibitem [{\citenamefont {\ifmmode \check{S}\else
  \v{S}\fi{}ari\ifmmode~\acute{c}\else \'{c}\fi{}}\ and\ \citenamefont
  {Cacciuto}(2012)}]{Cacciuto_2012a}%
  \BibitemOpen
  \bibfield  {author} {\bibinfo {author} {\bibfnamefont {A.}~\bibnamefont
  {\ifmmode \check{S}\else \v{S}\fi{}ari\ifmmode~\acute{c}\else \'{c}\fi{}}}\
  and\ \bibinfo {author} {\bibfnamefont {A.}~\bibnamefont {Cacciuto}},\ }\href
  {\doibase 10.1103/PhysRevLett.109.188101} {\bibfield  {journal} {\bibinfo
  {journal} {Phys. Rev. Lett.}\ }\textbf {\bibinfo {volume} {109}},\ \bibinfo
  {pages} {188101} (\bibinfo {year} {2012})}\BibitemShut {NoStop}%
\bibitem [{\citenamefont {Bahrami}\ \emph {et~al.}(2012)\citenamefont
  {Bahrami}, \citenamefont {Lipowsky},\ and\ \citenamefont
  {Weikl}}]{Bahrami_2012}%
  \BibitemOpen
  \bibfield  {author} {\bibinfo {author} {\bibfnamefont {A.~H.}\ \bibnamefont
  {Bahrami}}, \bibinfo {author} {\bibfnamefont {R.}~\bibnamefont {Lipowsky}}, \
  and\ \bibinfo {author} {\bibfnamefont {T.~R.}\ \bibnamefont {Weikl}},\ }\href
  {\doibase 10.1103/PhysRevLett.109.188102} {\bibfield  {journal} {\bibinfo
  {journal} {Phys. Rev. Lett.}\ }\textbf {\bibinfo {volume} {109}},\ \bibinfo
  {pages} {188102} (\bibinfo {year} {2012})}\BibitemShut {NoStop}%
\bibitem [{Note2()}]{Note2}%
  \BibitemOpen
  \bibinfo {note} {The wrapping fraction is defined as the ratio of the
  particle area adhered to the membrane, $A_{ad}$, to the total area of the
  particle, $A$. $A_{ad}/A=0 $ corresponds to the unwrapped state, while
  $A_{ad}/A \approx 1 $ indicates a fully-wrapped state.}\BibitemShut {Stop}%
\bibitem [{Note3()}]{Note3}%
  \BibitemOpen
  \bibinfo {note} {Unlike for prolate ellipsoids at a phase boundary dominated
  by interfacial tension \cite {lehle08}, and although the boundary between the
  adhered and the free membrane is allowed to adjust freely in early stages of
  the energy minimization, we find an elliptical contact line in Fig.~\ref
  {WEDPEOE}~(b) without significant variation in its height.}\BibitemShut
  {Stop}%
\bibitem [{Note4()}]{Note4}%
  \BibitemOpen
  \bibinfo {note} {Our numerical calculations and the fit function cannot
  capture a partially-wrapped at very high wrapping fraction, see Ref.~\protect
  \rev@citealp {deserno04}. Therefore the adhesion strength for that
  ${S_\protect \mathrm {21}}$ occurs is obtained from the condition that the
  slope of the energy is zero at complete wrapping.}\BibitemShut {Stop}%
\bibitem [{Note5()}]{Note5}%
  \BibitemOpen
  \bibinfo {note} {It should be a partially-wrapped state at high wrapping
  fraction \cite {deserno04}, but this cannot be described by the function in
  Eq.~(\ref {fiteq}).}\BibitemShut {Stop}%
\bibitem [{\citenamefont {Seifert}\ and\ \citenamefont
  {Lipowsky}(1990)}]{seifert1990}%
  \BibitemOpen
  \bibfield  {author} {\bibinfo {author} {\bibfnamefont {U.}~\bibnamefont
  {Seifert}}\ and\ \bibinfo {author} {\bibfnamefont {R.}~\bibnamefont
  {Lipowsky}},\ }\href@noop {} {\bibfield  {journal} {\bibinfo  {journal}
  {Phys. Rev. A}\ }\textbf {\bibinfo {volume} {42}},\ \bibinfo {pages} {4768}
  (\bibinfo {year} {1990})}\BibitemShut {NoStop}%
\bibitem [{\citenamefont {Yi}\ \emph {et~al.}(2011)\citenamefont {Yi},
  \citenamefont {Shi},\ and\ \citenamefont {Gao}}]{Gao11}%
  \BibitemOpen
  \bibfield  {author} {\bibinfo {author} {\bibfnamefont {X.}~\bibnamefont
  {Yi}}, \bibinfo {author} {\bibfnamefont {X.}~\bibnamefont {Shi}}, \ and\
  \bibinfo {author} {\bibfnamefont {H.}~\bibnamefont {Gao}},\ }\href@noop {}
  {\bibfield  {journal} {\bibinfo  {journal} {Phys. Rev. Lett.}\ }\textbf
  {\bibinfo {volume} {107}},\ \bibinfo {pages} {1} (\bibinfo {year}
  {2011})}\BibitemShut {NoStop}%
\bibitem [{\citenamefont {Sun}\ and\ \citenamefont
  {Wirtz}(2006)}]{Sun_Wirtz_2006}%
  \BibitemOpen
  \bibfield  {author} {\bibinfo {author} {\bibfnamefont {S.~X.}\ \bibnamefont
  {Sun}}\ and\ \bibinfo {author} {\bibfnamefont {D.}~\bibnamefont {Wirtz}},\
  }\href@noop {} {\bibfield  {journal} {\bibinfo  {journal} {Biophys. J.}\
  }\textbf {\bibinfo {volume} {90}},\ \bibinfo {pages} {L10} (\bibinfo {year}
  {2006})}\BibitemShut {NoStop}%
\bibitem [{\citenamefont {Zhang}\ and\ \citenamefont
  {Nguyen}(2008)}]{Zhang_Nguyen08}%
  \BibitemOpen
  \bibfield  {author} {\bibinfo {author} {\bibfnamefont {R.}~\bibnamefont
  {Zhang}}\ and\ \bibinfo {author} {\bibfnamefont {T.~T.}\ \bibnamefont
  {Nguyen}},\ }\href@noop {} {\bibfield  {journal} {\bibinfo  {journal} {Phys.
  Rev. E}\ }\textbf {\bibinfo {volume} {78}},\ \bibinfo {pages} {051903}
  (\bibinfo {year} {2008})}\BibitemShut {NoStop}%
\bibitem [{\citenamefont {Morris}\ and\ \citenamefont
  {Homann}(2001)}]{Morris_Homann_2001}%
  \BibitemOpen
  \bibfield  {author} {\bibinfo {author} {\bibfnamefont {C.~E.}\ \bibnamefont
  {Morris}}\ and\ \bibinfo {author} {\bibfnamefont {U.}~\bibnamefont
  {Homann}},\ }\href@noop {} {\bibfield  {journal} {\bibinfo  {journal} {J.
  Membr. Biol.}\ }\textbf {\bibinfo {volume} {179}},\ \bibinfo {pages} {79}
  (\bibinfo {year} {2001})}\BibitemShut {NoStop}%
\bibitem [{\citenamefont {Zhang}\ \emph {et~al.}(2009)\citenamefont {Zhang},
  \citenamefont {Li}, \citenamefont {Lykotrafitis}, \citenamefont {Bao},\ and\
  \citenamefont {Suresh}}]{Zhang_Suresh_2009}%
  \BibitemOpen
  \bibfield  {author} {\bibinfo {author} {\bibfnamefont {S.}~\bibnamefont
  {Zhang}}, \bibinfo {author} {\bibfnamefont {J.}~\bibnamefont {Li}}, \bibinfo
  {author} {\bibfnamefont {G.}~\bibnamefont {Lykotrafitis}}, \bibinfo {author}
  {\bibfnamefont {G.}~\bibnamefont {Bao}}, \ and\ \bibinfo {author}
  {\bibfnamefont {S.}~\bibnamefont {Suresh}},\ }\href@noop {} {\bibfield
  {journal} {\bibinfo  {journal} {Adv. mater.}\ }\textbf {\bibinfo {volume}
  {21}},\ \bibinfo {pages} {419} (\bibinfo {year} {2009})}\BibitemShut
  {NoStop}%
\bibitem [{\citenamefont {Gao}\ \emph {et~al.}(2005)\citenamefont {Gao},
  \citenamefont {Shi},\ and\ \citenamefont {Freund}}]{Gao05}%
  \BibitemOpen
  \bibfield  {author} {\bibinfo {author} {\bibfnamefont {H.}~\bibnamefont
  {Gao}}, \bibinfo {author} {\bibfnamefont {W.}~\bibnamefont {Shi}}, \ and\
  \bibinfo {author} {\bibfnamefont {L.~B.}\ \bibnamefont {Freund}},\
  }\href@noop {} {\bibfield  {journal} {\bibinfo  {journal} {Proc. Natl. Acad.
  Sci. U.S.A.}\ }\textbf {\bibinfo {volume} {102}},\ \bibinfo {pages} {9469}
  (\bibinfo {year} {2005})}\BibitemShut {NoStop}%
\bibitem [{\citenamefont {Chaudhuri}\ \emph {et~al.}(2011)\citenamefont
  {Chaudhuri}, \citenamefont {Battaglia},\ and\ \citenamefont
  {Golestanian}}]{chaudhuri11}%
  \BibitemOpen
  \bibfield  {author} {\bibinfo {author} {\bibfnamefont {A.}~\bibnamefont
  {Chaudhuri}}, \bibinfo {author} {\bibfnamefont {G.}~\bibnamefont
  {Battaglia}}, \ and\ \bibinfo {author} {\bibfnamefont {R.}~\bibnamefont
  {Golestanian}},\ }\href@noop {} {\bibfield  {journal} {\bibinfo  {journal}
  {Phys. Biol.}\ }\textbf {\bibinfo {volume} {8}},\ \bibinfo {pages} {9}
  (\bibinfo {year} {2011})}\BibitemShut {NoStop}%
\bibitem [{\citenamefont {Adriani}\ \emph {et~al.}(2012)\citenamefont
  {Adriani}, \citenamefont {De~Tullio}, \citenamefont {Ferrari}, \citenamefont
  {Hussain}, \citenamefont {Pascazio}, \citenamefont {Liu},\ and\ \citenamefont
  {Decuzzi}}]{Adriani_Decuzzi_2012}%
  \BibitemOpen
  \bibfield  {author} {\bibinfo {author} {\bibfnamefont {G.}~\bibnamefont
  {Adriani}}, \bibinfo {author} {\bibfnamefont {M.~D.}\ \bibnamefont
  {De~Tullio}}, \bibinfo {author} {\bibfnamefont {M.}~\bibnamefont {Ferrari}},
  \bibinfo {author} {\bibfnamefont {F.}~\bibnamefont {Hussain}}, \bibinfo
  {author} {\bibfnamefont {G.}~\bibnamefont {Pascazio}}, \bibinfo {author}
  {\bibfnamefont {X.}~\bibnamefont {Liu}}, \ and\ \bibinfo {author}
  {\bibfnamefont {P.}~\bibnamefont {Decuzzi}},\ }\href@noop {} {\bibfield
  {journal} {\bibinfo  {journal} {Biomaterials}\ }\textbf {\bibinfo {volume}
  {33}},\ \bibinfo {pages} {5504} (\bibinfo {year} {2012})}\BibitemShut
  {NoStop}%
\bibitem [{\citenamefont {Zhang}\ \emph
  {et~al.}(2012{\natexlab{b}})\citenamefont {Zhang}, \citenamefont {Tekobo},
  \citenamefont {Tu}, \citenamefont {Zhou}, \citenamefont {Jin}, \citenamefont
  {Dergunov}, \citenamefont {Pinkhassik},\ and\ \citenamefont
  {Yan}}]{Zhang_Yan_2012}%
  \BibitemOpen
  \bibfield  {author} {\bibinfo {author} {\bibfnamefont {Y.}~\bibnamefont
  {Zhang}}, \bibinfo {author} {\bibfnamefont {S.}~\bibnamefont {Tekobo}},
  \bibinfo {author} {\bibfnamefont {Y.}~\bibnamefont {Tu}}, \bibinfo {author}
  {\bibfnamefont {Q.}~\bibnamefont {Zhou}}, \bibinfo {author} {\bibfnamefont
  {X.}~\bibnamefont {Jin}}, \bibinfo {author} {\bibfnamefont {S.~A.}\
  \bibnamefont {Dergunov}}, \bibinfo {author} {\bibfnamefont {E.}~\bibnamefont
  {Pinkhassik}}, \ and\ \bibinfo {author} {\bibfnamefont {B.}~\bibnamefont
  {Yan}},\ }\href@noop {} {\bibfield  {journal} {\bibinfo  {journal} {ACS Appl.
  Mater. Interfaces}\ ,\ \bibinfo {pages} {4099}} (\bibinfo {year}
  {2012}{\natexlab{b}})}\BibitemShut {NoStop}%
\bibitem [{\citenamefont {Decuzzi}\ and\ \citenamefont
  {Ferrari}(2008)}]{Decuzzi_Ferrari_2008}%
  \BibitemOpen
  \bibfield  {author} {\bibinfo {author} {\bibfnamefont {P.}~\bibnamefont
  {Decuzzi}}\ and\ \bibinfo {author} {\bibfnamefont {M.}~\bibnamefont
  {Ferrari}},\ }\href@noop {} {\bibfield  {journal} {\bibinfo  {journal}
  {Biophys. J.}\ }\textbf {\bibinfo {volume} {94}},\ \bibinfo {pages} {3790}
  (\bibinfo {year} {2008})}\BibitemShut {NoStop}%
\end{thebibliography}%

\end{document}